\documentclass[twocolumn,twocolappendix]{aastex631}

\usepackage{mysymbol}
\usepackage{comment}

\shorttitle{Multiple Outflows in IRAS 15398}
\shortauthors{Sai et al.}

\begin{document}

\title{Multiple Outflows around a Single Protostar IRAS 15398$-$3359}

\author[0000-0003-4361-5577]{Jinshi Sai (Insa Choi)}
\affiliation{Academia Sinica Institute of Astronomy and Astrophysics, 11F of Astro-Math Bldg, 1, Sec. 4, Roosevelt Rd, Taipei 10617, Taiwan}

\author[0000-0003-1412-893X]{Hsi-Wei Yen}
\affiliation{Academia Sinica Institute of Astronomy and Astrophysics, 11F of Astro-Math Bldg, 1, Sec. 4, Roosevelt Rd, Taipei 10617, Taiwan}

\author[0000-0002-0963-0872]{Masahiro N. Machida}
\affiliation{Department of Earth and Planetary Sciences, Faculty of Sciences, Kyushu University, Fukuoka, Fukuoka 819-0395, Japan}

\author[0000-0003-0998-5064]{Nagayoshi Ohashi}
\affiliation{Academia Sinica Institute of Astronomy and Astrophysics, 11F of Astro-Math Bldg, 1, Sec. 4, Roosevelt Rd, Taipei 10617, Taiwan}

\author[0000-0002-8238-7709]{Yusuke Aso}
\affiliation{Korea Astronomy and Space Science Institute, 776 Daedeok-daero, Yuseong-gu, Daejeon 34055, Republic of Korea}

\author[0000-0002-3801-8754]{Ana\"{e}lle J. Maury}
\affiliation{AIM, CEA, CNRS, Universit\`{e} Paris-Saclay, Universit\`{e} Paris Diderot, Sorbonne Paris Cit\`{e}, 91191 Gif-sur-Yvette, France}
\affiliation{Harvard--Smithsonian Center for Astrophysics, Cambridge, MA02138, USA}

\author[0000-0003-1104-4554]{S\'{e}bastien Maret}
\affiliation{Univ. Grenoble Alpes, CNRS, IPAG, 38000 Grenoble, France}



\begin{abstract}
We present the results of our mosaic observations of a single Class 0 protostar IRAS 15398$-$3359 with Atacama Compact Array (ACA) in the CO $J=2$--1 line. The new observations covering a $\rtsim2'$ square region revealed elongated redshifted and blueshifted components, which are located at distances of $\rtsim30''$--$75''$ on the northern and southern sides of the protostar, respectively, in addition to the previously observed primary and secondary outflows. These elongated components exhibit Hubble-law like velocity structures, i.e., an increase of velocity with increasing distance from the protostar, suggesting that it is the third outflow associated with the protostar. Besides, a new redshifted component is detected at radii of $\rtsim40''$--$75''$ on the northwestern side of the protostar. This redshifted component also exhibits a Hubble-law like velocity profile, which could be the counterpart of the secondary outflow mostly detected at blueshifted velocities in a previous study. The three outflows are all misaligned by $\rtsim20$--$90^\circ$, and the dynamical timescale of the primary outflow is shorter than those of the other outflows approximately by an order of magnitude. These facts hint that the outflow launch direction has significantly changed with time. The outflow direction may change if the rotational axis and the magnetic field are misaligned, or if the dense core is turbulent. We favor the second scenario as the origin of the multiple outflows in IRAS 15398$-$3359 based on a comparison between the observational results and numerical simulations.
\end{abstract}



\section{Introduction} \label{sec:intro}
Low-mass star formation occurs in dense cores through gravitational collapse. Throughout the star formation process, circumstellar disks form around protostars as a consequence of conservation of the angular momentum inherent in the initial dense cores \citep{Terebey:1984aa}. Revealing the disk formation process is essential for understanding the diversity of protoplanetary disks and planets forming within them.

Early theoretical works developed analytical models of the disk formation process considering the gravitational collapse of a rotating dense core \citep[e.g.,][]{Terebey:1984aa, Basu:1998aa}. On the other hand, later ideal magnetohydrodynamics (MHD) simulations indicated that the presence of a magnetic field aligned with the rotational axis of the dense core prevents disks from forming, because the magnetic field extracts angular momentum too efficiently \citep[][]{Allen:2003aa, Mellon:2008aa}. These results contradicted the fact that protostellar/protoplanetary disks are ubiquitous around young stellar objects (YSOs) \citep[e.g.,][]{Williams:2011aa, Andrews:2020aa, Tobin:2020aa, Ohashi:2023aa}. This issue has been solved by more detailed numerical simulations. Turbulence in dense cores, and/or misalignment between the magnetic field and the rotational axis of the dense core can suppress the efficiency of magnetic braking and allow disks to form \citep[e.g.,][]{Joos:2012aa, Seifried:2013aa, Matsumoto:2017aa}. Non-ideal MHD effects also enable sizable disks to form even in strongly magnetized dense cores, as the magnetic field can be partially decoupled from neutral gas \citep{Machida:2011aa, Tomida:2015aa, Hennebelle:2020aa}.

The presence of both turbulence and misalignment between the rotational axis and the associated magnetic field in dense cores have been observed. Molecular line observations of dense cores have revealed complex velocity structures that are not explained by a simple rotational motion, suggesting the presence of turbulence in these dense cores \citep{Caselli:2002aa, Chen:2019ab, Gaudel:2020aa, Sai:2023aa}. Polarization observations in continuum emission have reported that magnetic fields are randomly aligned with outflows associated with protostars, which may represent the rotational axes of disks \citep{Hull:2014aa, Yen:2021aa}. Velocity gradients and magnetic fields of dense cores, and disk orientations are all frequently found to be misaligned \citep{Galametz:2018aa, Galametz:2020aa, Yen:2021ab, Gupta:2022aa}, and these misalignments might affect the efficiency of the angular momentum transfer from large to small scales \citep{Galametz:2020aa, Gupta:2022aa}.

Although dense cores may be commonly turbulent or associated with a misaligned magnetic field, the dynamical role of these conditions in the star formation process remains still unclear. Numerical simulations of disk formation in turbulent dense cores or those associated with misaligned magnetic fields predict variations in the direction of the disk normal over time, because the angular momentum vector of infalling material changes with time \citep{Seifried:2013aa, Matsumoto:2017aa, Bate:2018aa, Hirano:2019aa, Machida:2020aa}. In these simulations, the outflow orientation is also expected to vary with time as consequence, given that the outflow is typically launched in the direction of the disk normal \citep{Seifried:2013aa, Machida:2020aa}.

IRAS 15398$-$3359 (hereafter IRAS 15398) is a Class 0 protostar in the Lupus I star forming cloud at a distance of $155~\pc$ \citep{Zucker:2020aa, Santamaria-Miranda:2021aa,Thieme:2023aa}. Its bolometric temperature and luminosity are $50~\kelv$ and $1.4~\Lsun$, respectively \citep{Ohashi:2023aa}. The protostar is embedded in an infalling and rotating envelope and associated with a primary bipolar outflow along the northeast-southwest direction \citep{Oya:2014aa, Bjerkeli:2016ab, Yen:2017aa}. In addition to the primary outflow, a monopolar, blueshifted secondary outflow launched in a direction orthogonal to the primary outflow has been reported through molecular line observations with a high sensitivity using Atacama Large Millimeter/submillimeter Array (ALMA) \citep{Okoda:2021aa}. Molecular line observations with ALMA have also suggested the presence of a Keplerian disk with a radius of $\gtrsim30~\au$ \citep{Okoda:2018aa, Thieme:2023aa}. Previous works have inferred a very small protostellar mass of $\lesssim 0.01~\Msun$ based on observations at angular resolutions of $\lesssim 0\farcs2$ \citep{Yen:2017aa, Okoda:2018aa}. Recent observations at an angular resolution of $\rtsim0\farcs15$ have revealed a possible differential rotation signature of a Keplerian disk, and constrained the central protostellar mass range of $\rtsim0.02$--$0.1~\Msun$ through fitting to the differential rotation signature \citep{Thieme:2023aa}.

In this paper, we report the discovery of a third outflow around IRAS 15398, which is misaligned with the known primary and secondary outflows, and discuss the possible origins of the misaligned, multiple outflows. The finding of the largely misaligned multiple outflows in IRAS 15398 hints the dynamical importance of turbulence in the dense core and/or misalignment between the magnetic field and the rotational axis of the dense core. The outline of this paper is as follows. Observations and data reduction are summarized in Section \ref{sec:obs}. Observational results are presented in Section \ref{sec:res}. Analysis on velocity structures is provided in Section \ref{sec:ana}, and possible implications are discussed in Section \ref{sec:disc}. Main results and conclusions are summarized in Section \ref{sec:summary}.

\section{ACA 7-m array Observations} \label{sec:obs}
We have conducted mosaic observations of IRAS 15398 using the 7-m array of Atacama Compact Array (ACA) on 2019 December 19 during ALMA Cycle 7 with 10 antennae in Band 6. In this paper, we present the CO $J=$2--1 (230.538000 GHz) results of the observations. The mosaic observations with 28 pointings covered $\rtsim2'$ square region centered at the protostellar position. The projected baseline lengths ranged from 6.5 to 37 k$\lambda$ at 230 GHz. The observations were carried out in the frequency division mode (FDM). A spectral window with a bandwidth of 62.5 MHz and a spectral resolution of 61.0 kHz was assigned to the CO $J=$2--1 line, which provided a velocity resolution of $\sim$0.080 $\kmps$. J1534$-$3526 and J1337$-$1257 were observed for the phase calibration and for the bandpass and flux calibrations, respectively. The data were calibrated with the pipeline version 42866M (Pipeline-CASA56-P1-B) of the Common Astronomy Software Applications package \citep[CASA;][]{McMullin:2007aa} version 5.6.1.

We produced images from the visibility data using CASA 5.6.1 with the \texttt{tclean} task. We adopted the Briggs style weighting with the robust parameter of 0.5 and the velocity resolution of $0.08~\kmps$. Multiscale clean was used with the scale parameters of zero and 10, which correspond to the point source and about one beam size, respectively. The primary beam correction was applied to the cleaned image. We produced a final image cube with a velocity resolution of 0.16 $\kmps$ for analysis by binning two velocity channels. The angular resolution and root-mean-square (rms) noise level are $7\farcs7\times 4\farcs1$ (87$^\circ$) and $0.13~\jypbm$, respectively. We adopted the source position of ($\alpha_\mathrm{J2000}$, $\delta_\mathrm{J2000}$)$=$(15:43:02.24, $-34$:09:06.8) for the figures and analyses in this paper \citep{Yen:2017aa}.

\section{Results} \label{sec:res}
\begin{figure*}[thbp]
\plotone{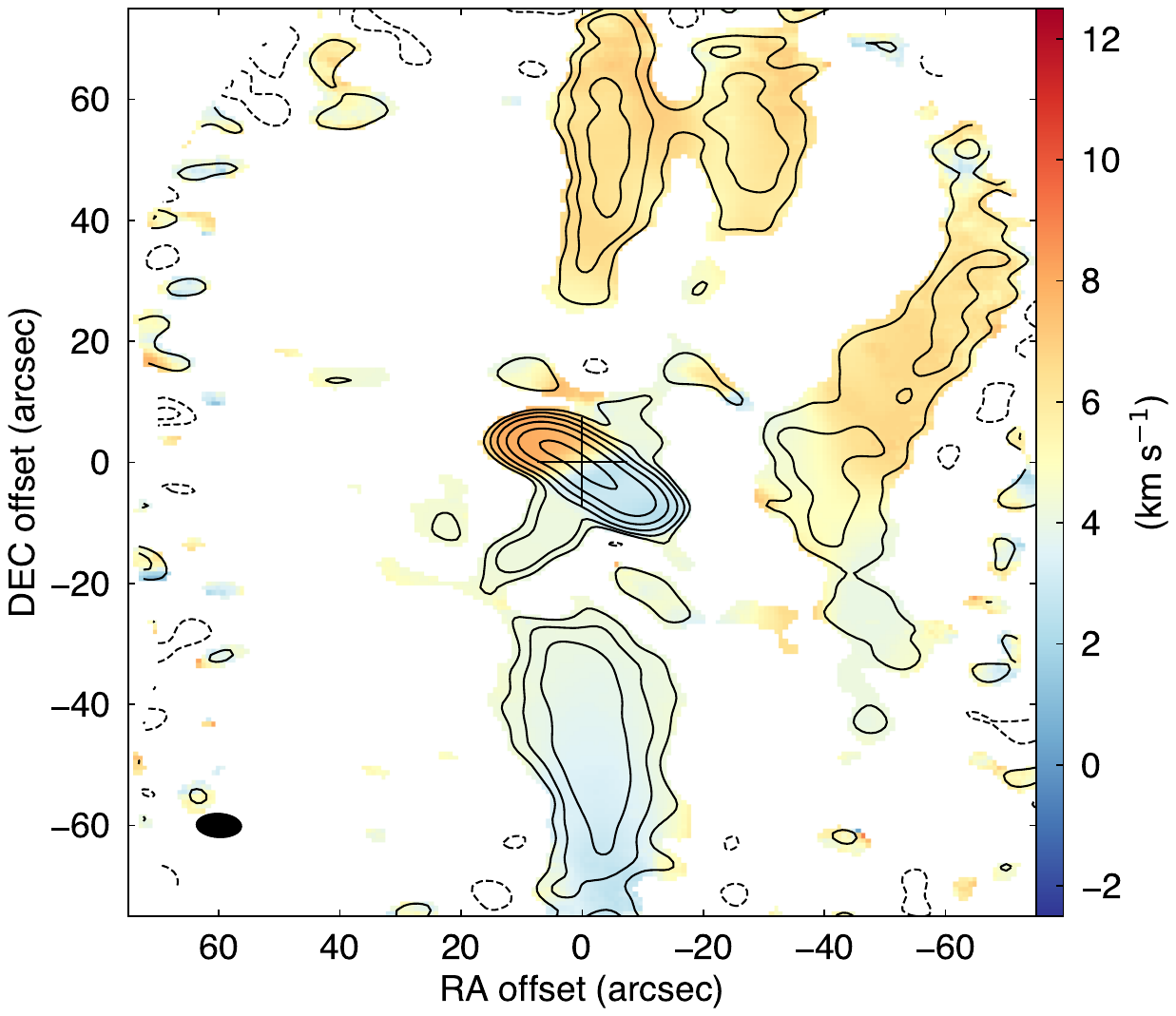}
\caption{Moment 0 (contours) and 1 (color) maps of the CO $J=2$--1 emission of IRAS 15398$-$3359. Contour levels are 3, 6, 12, 24, ... $\times\sigma$, where 1$\sigma$ is 0.58 $\jypbm$. The central cross denotes the protostellar position. The ellipse at the bottom left corner indicates the synthesized beam size of $7\farcs7\times 4\farcs1$ (87$^\circ$). \label{fig:momentmaps}}
\end{figure*}

Moment 0 and 1 maps of the CO $J=2$--1 emission are presented in Figure \ref{fig:momentmaps}. The emission reveals several elongated structures at various spatial scales. Within a radius of $30''$ around the protostellar position, we see an elongated structure from southwest to northeast in a length of $\sim$30$''$. The elongated emission shows a velocity gradient in the same direction: a blueshifted component on the southwestern side and a redshifted component on the northeastern side of the protostar. Both blueshifted and redshifted components have a relatively high velocity of $\pm$3 $\kmps$ with respect to the systemic velocity of 5.4 $\kmps$ \citep{Thieme:2023aa} in the moment 1 map. This elongated structure likely traces the primary outflow that has been reported in previous observations \citep{Oya:2014aa, Yen:2017aa, Okoda:2021aa} based on its morphology and velocity structure. Another component extends from the protostellar position to the southeast. This component has a LSR velocity of $\rtsim4$--5 $\kmps$. Its elongation direction is close to those of the major axis of the disk and envelope \citep{Yen:2017aa, Okoda:2018aa, Thieme:2023aa} and the monopolar, blueshifted secondary outflow \citep[P.A.$\rtsim140^\circ$; ][]{Okoda:2021aa}, although its size is larger than the size of the secondary outflow. At distances from the protostar larger than $30''$, blueshifted and redshifted components are observed on the southern and northern sides of the protostar, respectively. These components appear symmetric with respect to the protostellar position. They also exhibit velocity gradients in the north-south direction with relative velocities of $\sim$1--2 $\kmps$ with respect to the systemic velocity. These features suggest that they are an bipolar outflow associated with IRAS 15398.
Another elongated, redshifted emission is found at $(\Delta \alpha, \Delta \delta)\tsim(-30'', 50'')$, which is approximately aligned with a line passing through the protostellar position and the blueshifted component elongated in the southeast-northwest direction. Another elongated component is located at $\sim$40$''$ west from the protostellar position.

\begin{figure*}
\plotone{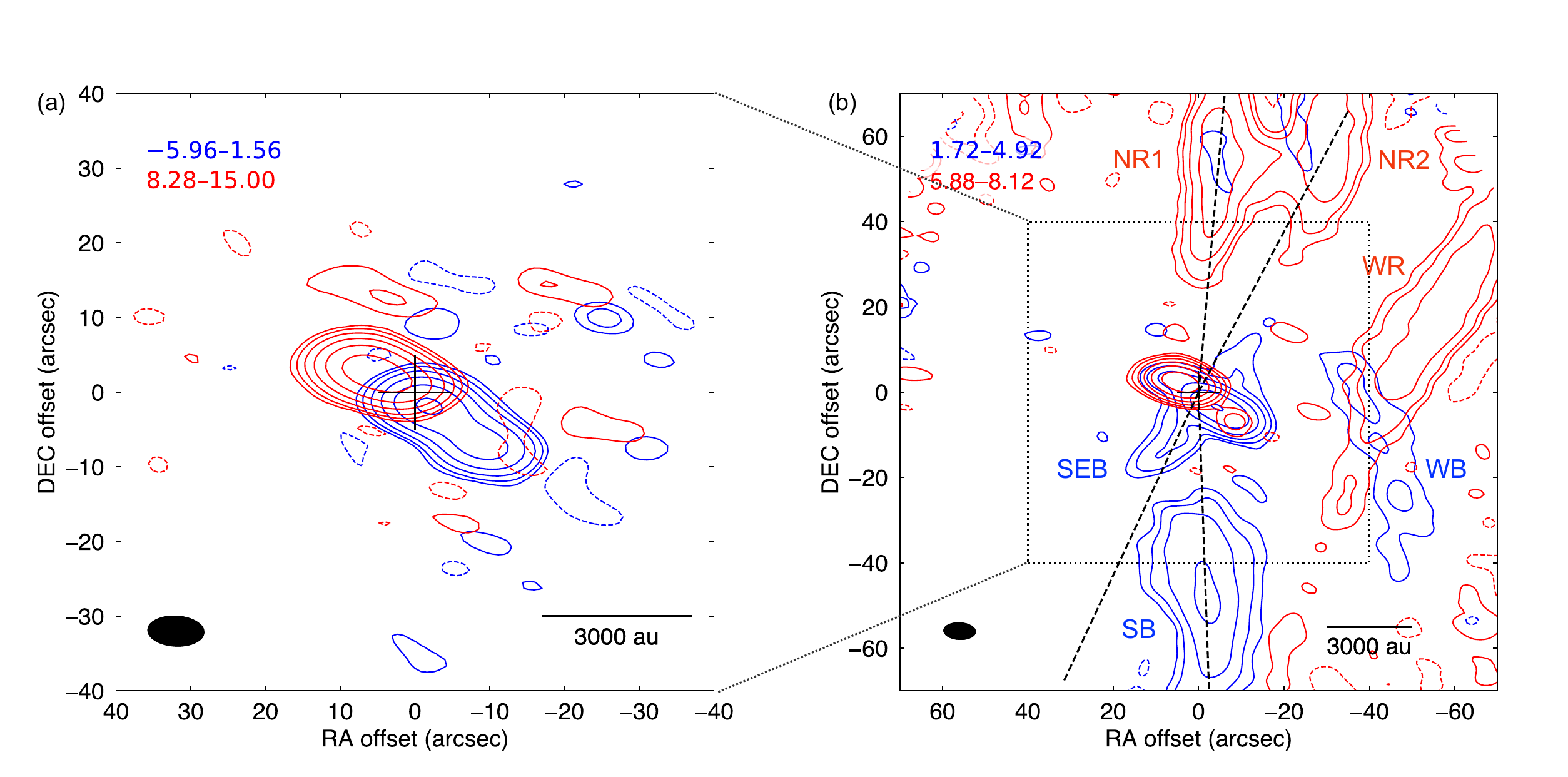}
\caption{(a) High- and (b) low-velocity components of the CO $J=$2--1 emission. Red and blue colors show redshifted and blueshifted components, respectively. The LSR velocity range of each velocity component is shown in the top left corners in units of $\kmps$. Contour levels are 3, 6, 12, 24, ... $\times\sigma$, where 1$\sigma$ is the rms noise level. The rms of each map is 0.16 $\jypbm$ for the high-velocity redshifted and high-velocity blueshifted components, 0.20 $\jypbm$ for the low-velocity redshifted component, and 0.46 $\jypbm$ for the low-velocity blueshifted component. The central crosses denote the protostellar position. The ellipses at the bottom left corners indicate the synthesized beam size of $7\farcs7\times 4\farcs1$ (87$^\circ$). The dashed lines in panel b show directions of the position-velocity (PV) cuts. \label{fig:maps_diffvel}}
\end{figure*}
%

In order to probe further details of the velocity structures of the CO emission, moment 0 maps of different velocity components are presented in Figure \ref{fig:maps_diffvel} (see also velocity channel maps shown in Figure \ref{fig:channels}). The high-velocity components shown in Figure \ref{fig:maps_diffvel}(a) correspond to the emission elongated from the southwest to northeast near the protostar in Figure \ref{fig:momentmaps}. These components are detected over a wide velocity range: the high-velocity blueshifted and redshifted components range from $-6.0$ to $1.6~\kmps$ and from 8.3 to $15.0~\kmps$, respectively. More components appear at lower velocities, as shown in Figure \ref{fig:maps_diffvel}(b). The northern redshifted (NR1 and NR2) components and the southern blueshifted (SB) component have velocities of $\rtsim5.9\mbox{--}8.1~\kmps$ and $\rtsim1.7\mbox{--}4.9~\kmps$, respectively. The blueshifted elongation from the protostellar position toward the southeast (SEB) and the blueshifted filamentary structure on the western side of the protostar (WB) have LSR velocities of $\rtsim3.5\mbox{--}4.8~\kmps$, which is closer to the systemic velocity of $5.4~\kmps$ than the velocity of the SB component (see Figure \ref{fig:channels}). The redshifted filamentary structure on the western side (WR) also has a smaller velocity of $\rtsim6.4$--$7.1~\kmps$ than the the NR1 and NR2 components.

\section{Analysis} \label{sec:ana}
\subsection{Velocity Structure} \label{subsec:ana_vfield}
The velocity structures of low-velocity components are investigated in more detail in this section.
\begin{figure*}
\epsscale{1.2}
\plotone{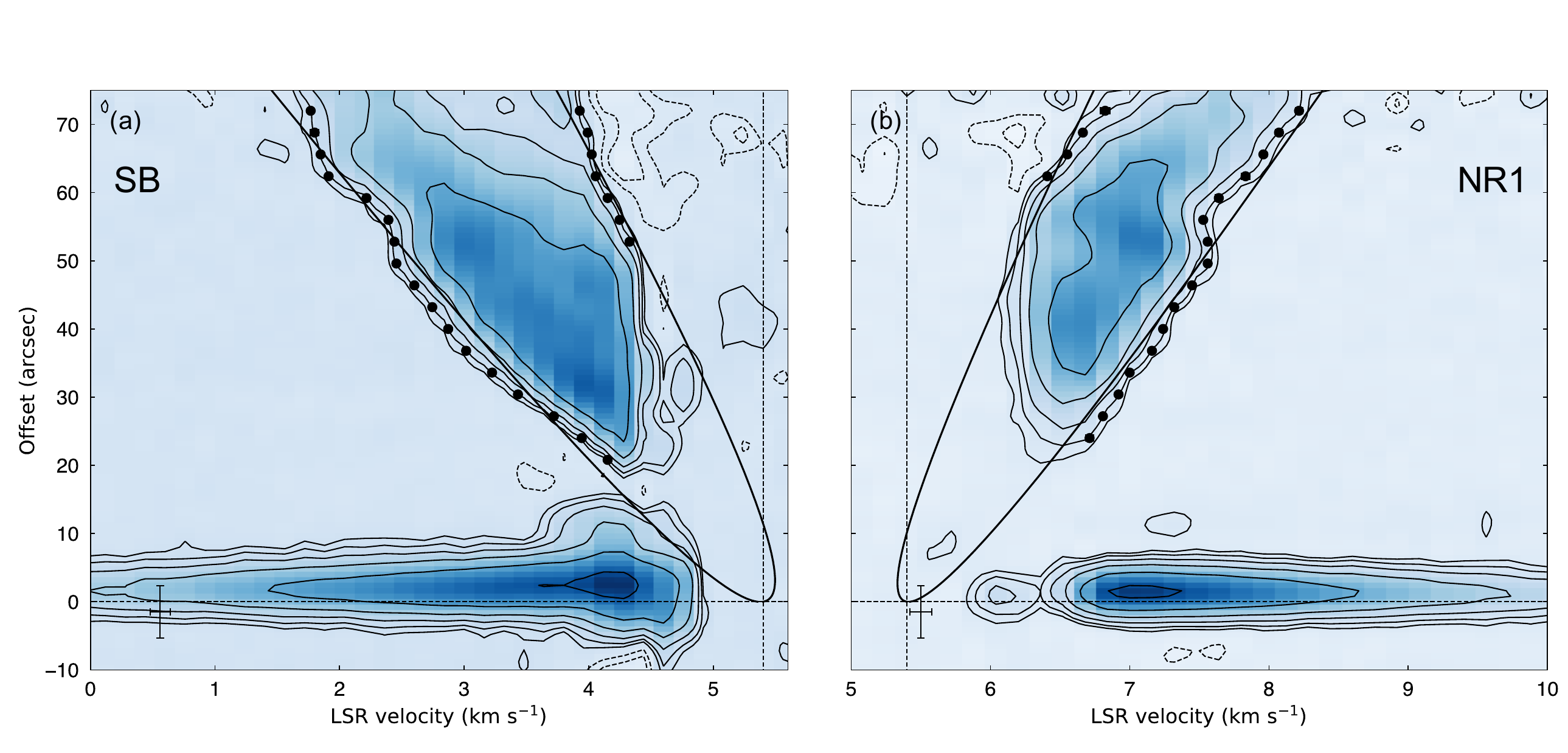}
\caption{PV diagrams of the CO $J=$2--1 emission cut toward the (a) SB and (b) NR1 components. Contour levels are from 3, 6, 12, 24, ... $\times\sigma$, where 1$\sigma$ is 0.13 $\jypbm$. Black dots are data points tracing $5\sigma$ contours that are used to fit the wind-driven shell model. Most of error bars of the data points are smaller than the marker size. Black curves show velocity structures of the best-fit wind-driven shell models. The vertical and horizontal bars at the bottom left corners indicate the synthesized beam size of $7\farcs7\times 4\farcs1$ ($87^\circ$) and the velocity resolution of $0.16~\kmps$. The vertical and horizontal dashed lines denote the systemic velocity of $5.4~\kmps$ and the protostellar position, respectively. \label{fig:pvds-3rdoutflow}}
\end{figure*}
Figure \ref{fig:pvds-3rdoutflow} presents position-velocity (PV) diagrams cut toward the NR1 (P.A.$=355^\circ$) and SB (P.A.$=182^\circ$) components. The position angle of each component is measured as an angle from north to the line passing through the protostellar position and the peak position of the emission, which is determined by a Gaussian fit. No clear emission is found around the systemic velocity of $5.4~\kmps$. This would be because the optically thick, extended foreground CO emission is resolved out. Two distinct components are found in each PV diagram. Within an offset of $10''$, the emission extends over velocity ranges of 0--$5~\kmps$ and 6--$10~\kmps$, which is the high-velocity components of the primary outflow. The NR1 and SB components are found at offsets of 20--$75''$. They exhibit the so-called Hubble-law like velocity profiles, i.e., an increase in the relative velocity with respect to the systemic velocity with increasing distance from the protostar. The velocity ranges of the NR1 and SB components are $\rtsim1$--$3.5~\kmps$ relative to the systemic velocity. These velocity structures and symmetric geometry with respect to the protostellar position suggest that the NR1 and SB components trace an outflow launched from IRAS 15398. This third outflow is highly misaligned with the primary and secondary outflows, as discussed in more detail later.

\begin{figure}
\epsscale{1.2}
\plotone{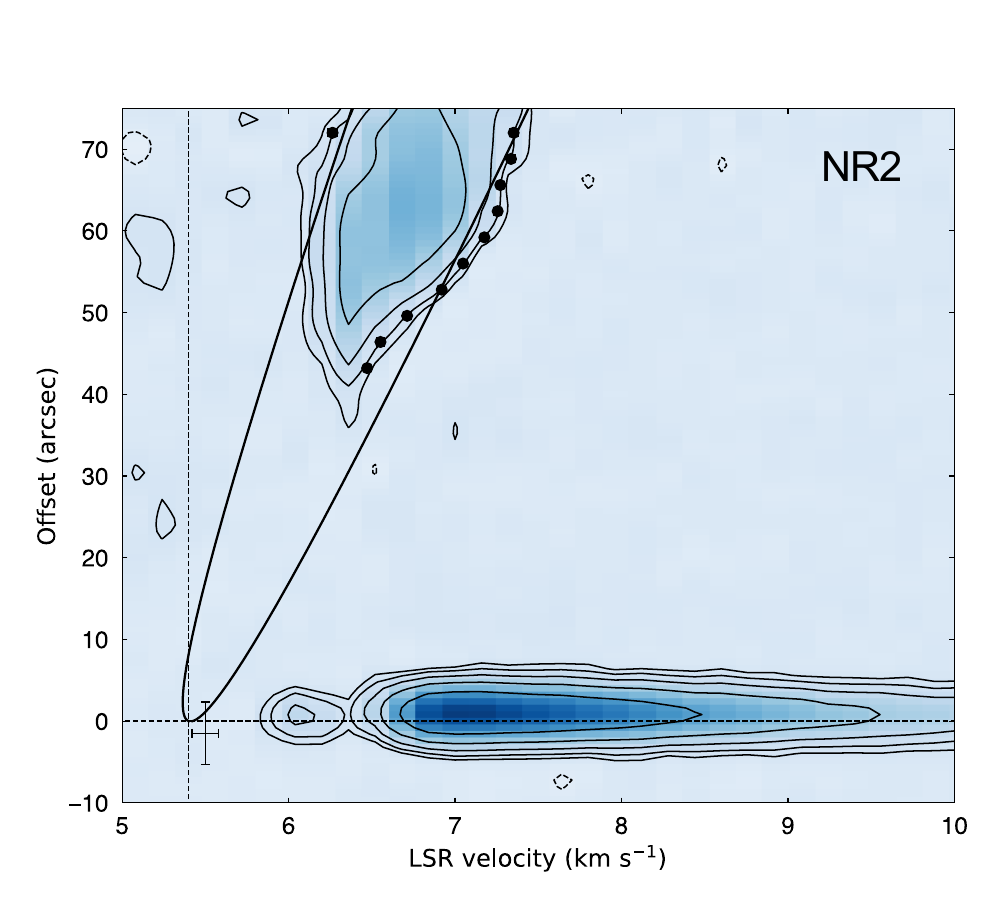}
\caption{PV diagram of the CO $J=$2--1 emission cut toward the NR2 component. Contour levels and symbols are the same as those in Figure \ref{fig:pvds-3rdoutflow}. \label{fig:pvd-nr2}}
\end{figure}

\begin{figure}
\epsscale{1.2}
\plotone{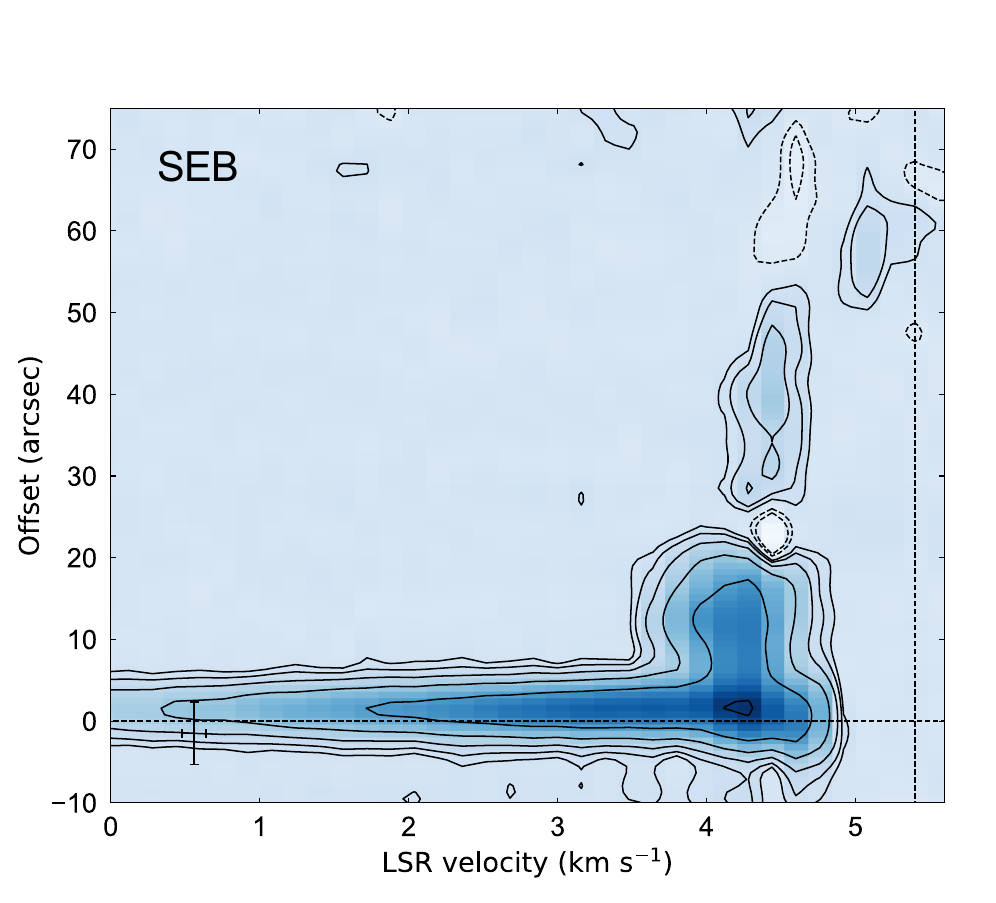}
\caption{PV diagram of the CO $J=$2--1 emission cut along the SEB component. Contour levels and symbols are the same as those in Figure \ref{fig:pvds-3rdoutflow}. \label{fig:pvd-2ndoutflow}}
\end{figure}

A PV diagram cut toward the NR2 component (P.A.$=332^\circ$) is presented in Figure \ref{fig:pvd-nr2}. The emission of the NR2 component is located at offsets of $40''$--$75''$, while the emission within an offset of $10''$ is associated with the primary outflow. The velocity structure of the NR2 component is similar to that of the NR1 component; the velocity increases from $\rtsim6$ to $7.5~\kmps$ as the offset increases. The Hubble-law like velocity profile suggests that the NR2 component is also part of an outflow. Although its symmetric counterpart is not found, the NR2 and the SEB components are almost aligned in the same line passing through the central protostar, as shown in Figure \ref{fig:maps_diffvel}(b). The relationship of the NR2 component to the other outflows is discussed in more detail in Section \ref{subsec:outflow_config}.

Figure \ref{fig:pvd-2ndoutflow} shows a PV diagram cut toward the SEB component (P.A.$=155^\circ$). While the emission within an offsets of $\rtsim10''$ is the high-velocity component of the primary outflow, the SEB component appears at offsets of $\rtsim10''$--$20''$ and exhibits an almost constant velocity of $\rtsim4.3~\kmps$. Its velocity structures are not well resolved with our limited spatial resolution. The other components at further larger offsets of $\sim$25$''$--70$''$ at velocities of $\rtsim$4.5 and $5~\kmps$ are likely part of ambient gas, as the CO emission is highly complex or almost resolved out at these velocities (see Figure \ref{fig:channels}). The SEB component is approximately along the disk/envelope major axis, as well as the secondary outflow detected on a smaller scale in previous observations \citep[][]{Yen:2017aa, Okoda:2018aa, Okoda:2021aa, Thieme:2023aa}. 
The previous studies have indicated that the infalling and rotating envelope exhibits redshifted velocity on the southeastern side, which does not match the blueshifted velocity of the SEB component. 
Thus, the SEB component is likely unrelated to the envelope. On the other hand, the velocity of the SEB component is similar to those of shock remnant cause by the secondary outflow \citep{Okoda:2021aa}. The SEB component and the secondary outflow are compared in more detail in Section \ref{subsec:outflow_config}.

\begin{figure*}
\epsscale{1.2}
\centering
\plotone{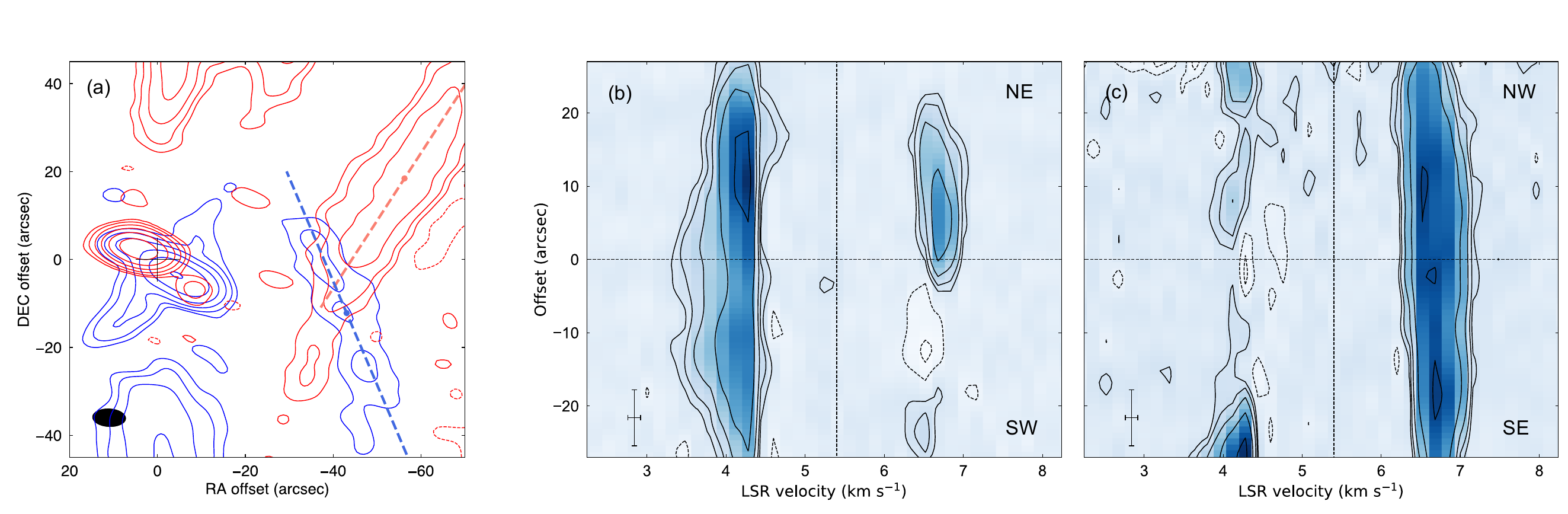}
\caption{(a) Same as Figure \ref{fig:maps_diffvel}(b) but zoomed into the WB and WR filamentary structures. Blue/red dashed lines and dots denote the directions and centers of the PV cuts, respectively. (b) PV diagrams of the CO 2--1 emission cut along the WB and (c) WR filamentary structures centered their emission peaks. The vertical and horizontal bars at the bottom left corners and contour levels are the same as those in Figure \ref{fig:pvds-3rdoutflow}. The vertical and horizontal dashed lines denote the systemic velocity of $5.4~\kmps$ and the emission peak of the filamentary structures. \label{fig:pvds-ambinet}}
\end{figure*}
The WB and WR filamentary structures do not appear to be pointing towards the protostellar position, which suggests that they may not be part of the outflows launched from IRAS 15398. The PV diagrams cut along the WB and WR filamentary structures are presented in Figure \ref{fig:pvds-ambinet}. Their centers and position angles are measured though Gaussian fitting and are presented in Figure \ref{fig:pvds-ambinet}(a). Figure \ref{fig:pvds-ambinet}(b) and (c) show that these structures have almost constant LSR velocities around 4.3 and $6.8~\kmps$, respectively, along the elongation. No emission is detected from $\rtsim4.5$ to $6.5~\kmps$ in the PV diagrams, which would be due to the resolved-out, optically thick foreground gas. The constant velocities close to the cloud velocity suggest that the WB and WR filamentary structures are part of the ambient cloud. Similar filamentary structures tracing large-scale ambient material have also been reported in other protostellar systems \citep{Dutta:2022aa}.

\subsection{Physical Properties of the Outflows}
The NR1 and SB components are likely a pair of lobes of the third outflow because of their symmetric morphology and velocity structures with the Hubble-law like velocity profiles. The NR2 component shows the Hubble-law like velocity profile and thus would also trace an outflow lobe. To characterize their geometry and velocity structures, we perform fitting with a wind-driven shell model \citep[][]{Shu:1991aa, Lee:2000aa}. In the wind-driven shell model, the morphology and velocity structure of the outflow cavity are expressed as follows:
\begin{eqnarray}
    z = C R^2,~~~v_\mathrm{R} = v_0\frac{R}{R_0},~~~v_\mathrm{z} = v_0\frac{z}{z_0},
\end{eqnarray}
where $R$ is the cylindrical radius in a direction perpendicular to the outflow axis, $z$ is the distance from the protostar along the outflow axis, and $v_R$ and $v_z$ are the expansion velocities. The constants $R_0$ and $z_0$ are fixed at $1''$, and the unit of $C$ is $\mathrm{arcsec}^{-1}$. The outflow cavity is then projected onto the plane of the sky with the inclination angle $i$. Here, we define the inclination angle as an angle between the line-of-sight and the outflow axis; in our definition, $i=90^\circ$ corresponds to the edge-on disk configuration. We search for the parameter set of ($C$, $v_0$, $i$) that best explains the observations. The fitting is performed in two steps. First, we fit the outflow shell morphology to constrain $C\sin i$ using the moment 0 maps presented in Figure \ref{fig:maps_diffvel}. Then, adopting the constrained $C\sin i$, we fit the velocity structure using the PV diagrams. In both steps, $5\sigma$ contours of the emission are used to define morphologies and velocity structures of the observed outflows. Details of the fitting process are presented in Appendix \ref{app:modelfit}.

The fitting results are presented in Figures \ref{fig:pvds-3rdoutflow}, \ref{fig:pvd-nr2} and \ref{fig:app_cavityfit} and summarized in Table \ref{tab:res-modelfit}. The inclination angles of the three lobes are all $\rtsim70^\circ$. While the expansion velocities ($v_0$) of the SB and NR1 lobes are very similar, the expansion velocity of the NR2 lobe is smaller, approximately by a factor of two, than those of the SB and NR1 lobes. This suggests that the NR2 lobe is tracing a different ejection event than the third outflow.

\begin{deluxetable}{lccc}
\tablecaption{Results of Wind-driven Shell Model Fitting} \label{tab:res-modelfit}
\tablehead{
\colhead{Lobes} & \colhead{$C \sin{i}$} & \colhead{$v_0$} & \colhead{$i$} \\
\colhead{} & \colhead{} & \colhead{($\kmps$)} & \colhead{($^\circ$)}}
\startdata
SB  & $0.81\pm0.02$ & $0.1129\pm0.0004$ & $71.83 \pm 0.05$ \\
NR1 & $1.24 \pm 0.02$ & $0.1017\pm0.0006$ & $74.24 \pm 0.12$ \\
NR2 & $0.89\pm 0.02$ & $0.0542\pm0.0016$ & $69.75 \pm 0.71$
\enddata
\end{deluxetable}


\begin{deluxetable}{lccccc}
\tablecaption{Lengths, Velocities and Dynamical Timescales of the Outflow Lobes} \label{tab:tdyn}
\tablehead{
\colhead{Lobes} & \colhead{$r_\mathrm{CO}$} & \colhead{$v_\mathrm{max,LSR}$} & \colhead{$v_\mathrm{max,rel}$} & \colhead{$\tau_\mathrm{dyn}'$} & \colhead{$\tau_\mathrm{dyn}$} \\
\colhead{} & \colhead{($\au$)} & \colhead{($\kmps$)} & \colhead{($\kmps$)} & \colhead{($10^4~\mathrm{yr}$)}  & \colhead{($10^4~\mathrm{yr}$)} 
}
\startdata
SB & $\gtrsim12,000$ & 1.72 & 3.68 & $\gtrsim1.5$ & $\gtrsim0.49$\\
NR1 & $\gtrsim12,000$ & 8.12 & 2.72 & $\gtrsim2.1$ & $\gtrsim0.59$ \\
NR2 & $\gtrsim12,000$ & 7.32 & 1.92 & $\gtrsim3.0$ & $\gtrsim1.1$
\enddata
\tablecomments{$\tau_\mathrm{dyn}$ and $\tau_\mathrm{dyn}'$ are the dynamical timescales with and without inclination-angle correction, respectively, and $\tau_\mathrm{dyn} = \tau_\mathrm{dyn}'/\tan i$.}
\end{deluxetable}
We also estimate the dynamical timescales of these lobes to probe their relationship with the known primary and secondary outflows. We use their lengths and the maximum velocities to calculate the dynamical timescales, following the method used in previous works \citep{Yildiz:2015aa, Okoda:2021aa}. The lengths of the outflows, their maximum velocities, and the derived dynamical timescales are summarized in Table \ref{tab:tdyn}. Note that all these lobes are extended to the map edge, and thus, the outflow lengths represent lower limits. Consequently, the calculated dynamical timescales are also lower limits. The dynamical timescales of the NR1, NR2 and SB lobes are calculated to be $\gtrsim0.49$--$1.1\times 10^4~\yr$ with correction of the inclination angle. A previous work estimated the dynamical timescale of the primary outflow to be $\rtsim 5 \times 10^2~\yr$ with correction of the inclination angle in the same manner \citep{Bjerkeli:2016ab}, suggesting that the SB, NR1 and NR2 lobes are launched first, and then followed by the primary outflow.

\section{Discussion} \label{sec:disc}
\subsection{Configurations of the Multiple Outflows}\label{subsec:outflow_config}

\begin{deluxetable}{cccccc}
\tablecaption{Position Angles of the Outflows} \label{tab:pa_outflow}
\tablehead{
\multicolumn{2}{c}{Primary} & \multicolumn{2}{c}{Secondary} & \multicolumn{2}{c}{Third} \\
\colhead{Red$^1$} & \colhead{Blue$^1$} & \colhead{Red$^3$} & \colhead{Blue$^2$} & \colhead{Red$^3$} & \colhead{Blue$^3$}  
}
\startdata
$60$ & $230$ & $332$ & $140$ & $355$ & $182$
\enddata
\tablecomments{$^1$\cite{Yen:2017aa}; $^2$\cite{Okoda:2021aa}; $^3$This work. All values are in units of degree.}
\end{deluxetable}
The SEB lobe is well aligned with the secondary outflow on a smaller scale of $\rtsim200$--$1200~\au$ \citep{Okoda:2021aa}. The previous work has also revealed an arc-like structure in SO and SiO lines at the tip of the secondary outflow at a distance of $1200~\au$ from the protostar. Based on the relatively narrow velocity range of the detected SO and SiO emission from $\rtsim4$ to $5.5~\kmps$, the authors suggested that the arc-like structure traces an old shock caused by the secondary outflow. The velocity range of the arc-like structure is almost consistent with the velocity range of the SEB lobe. The lack of the emission of the SEB lobe at velocities of 5--$5.5~\kmps$ would be due to the resolved-out, optically thick foreground gas (Figure \ref{fig:channels}). Faint, blueshifted CO emission was also detected near the SO/SiO arc-like structure in other interferometric observations \citep{Bjerkeli:2016ab, Yen:2017aa}. \cite{Okoda:2021aa} have pointed out that the faint CO emission would be a remnant of part of the arc-like structure. The location and velocity of the CO emission detected in the previous works are mostly coincident with those of the SEB lobe, although the SEB lobe also traces emission on larger scales than the previous observations. Thus, the SEB lobe could be a dissipating remnant of the secondary outflow.

The NR2 lobe appears aligned with the direction of the SEB lobe and the secondary outflow, suggesting that the NR2 lobe is the counterpart of the secondary outflow. The similar velocities of the NR2 and SEB lobes of $\rtsim1$--$2~\kmps$ with respect to the systemic velocity also support this possibility. Numerical simulations suggest that the outflow can be asymmetric in the presence of turbulence because of the interaction between the outflow and an inhomogeneous envelope \citep{Offner:2011aa}. \cite{Offner:2011aa} show that the lengths of two lobes of an outflow with velocities of a few $\kmps$ can differ by more than a few thousands au. One might wonder whether the NR1 and NR2 lobes originate from a cavity wall of a single outflow. However, their distinct expansion velocities suggest that they trace different ejection events, as was mentioned in Section \ref{subsec:ana_vfield}. Hence, the NR1 and NR2 lobes are most likely two different outflows but not different sides of the same outflow cavity. The dynamical timescale of the secondary outflow is estimated to be $\rtsim5\times 10^3~\yr$ without the correction of the inclination angle \citep{Okoda:2021aa}, which is shorter than that of the NR2 lobe. This would be because the blueshifted lobe of the secondary outflow seems to have experienced shock \citep{Okoda:2021aa} and has not propagated to radii larger than $\rtsim3000~\au$. Since the dynamical timescales for the SB, NR1 and NR2 are only lower limits, it is challenging to further discuss which outflow, the second or the third, is ejected earlier or later with the current datasets.

The configurations of the primary, secondary and third outflows on the plane of the sky are summarized in Figure \ref{fig:summary_config} and Table \ref{tab:pa_outflow}. Considering the position angles of the redshifted and blueshifted lobes of the outflows, the misalignment angles between these three outflows on the plane of the sky are $\rtsim20$--$90^\circ$.

\subsection{Origin of the Multiple Outflows}
\label{subsec:origins}
As discussed in \cite{Okoda:2021aa}, IRAS 15398 has been known as a single protostar so far. \cite{Ohashi:2023aa} and \cite{Thieme:2023aa} have recently presented 1.3 mm continuum images at an angular resolution of $0\farcs04$ ($\sim$6 au), indicating that the source is not a binary with a separation wider than $\rtsim6$ au. The presence of a Keplerian disk with a radius of 30 au around IRAS 15398 have been suggested \citep{Okoda:2018aa, Thieme:2023aa}. Observations of close binary systems show that most of them having a circumbinary disk with such a small binary separation exhibit a single outflow or parallel outflows \citep[][]{Lee:2015aa, Tobin:2016aa, Lim:2016aa, Alves:2017aa}, as pointed out by \cite{Okoda:2021aa}. Thus, these multiple outflows would not be likely driven by unresolved very close binary protostars.

Another possible explanation of the multiple outflows around IRAS 15398 is time variation of the outflow orientation. The orientation of the outflow is expected to vary with time when the angular momentum axis of the infalling material is randomly aligned. Indeed, the dynamical timescales of the outflows are different by an order of magnitude, as discussed above. Theoretical works predict that such time variation of the outflow orientation occurs when dense cores are turbulent or the associated magnetic field is misaligned with the initial rotational axis of the dense core \citep{Matsumoto:2004aa,Matsumoto:2017aa, Machida:2020aa}. In these simulations, the orientation of the disk normal changes with time because angular momentum vectors of infalling gas elements are not aligned with a single direction. Consequently, the outflow orientation also changes with time, since the outflow is generally launched in the direction of the disk normal. In the following subsections, whether the turbulence and misalignment scenarios fit IRAS 15398 will be discussed one by one.

\subsubsection{Turbulence Scenario} \label{subsec:turb_scenario}
Numerical simulations show that the disk orientation, which would determine the outflow direction, varies over time significantly, sometimes by more than $180^\circ$, in turbulent dense cores \citep{Bate:2010aa, Seifried:2013aa, Offner:2017aa, Matsumoto:2017aa, Bate:2018aa}. In turbulent dense cores, material infalling onto the disk has a random angular momentum vector and infalling material having larger angular momentum alters the net angular momentum vector of the disk. Some of these simulations yield that the disk orientation can change with time in dense cores when the ratio of the turbulent energy to the gravitational energy ($\beturb$) is $\rtsim0.06$--$0.09$ \citep{Seifried:2013aa, Matsumoto:2017aa}.

For a quantitative comparison to those simulations, we calculate $\beturb$ in the dense core of IRAS 15398 assuming the uniform density distribution:
\begin{equation}
\beturb =  \frac{E_\mathrm{turb}}{E_\mathrm{grav}},
\end{equation}
where
\begin{equation}
   \Eturb = \frac{3}{2} \Mcore \delta v_\mathrm{turb}^2,
\end{equation}
\begin{equation}
   \Egrav = \frac{3}{5} \frac{G \Mcore^2}{\Rcore}.
\end{equation}
Here, $\Eturb$ is the turbulent energy, $\Egrav$ is the gravitational energy, $\Mcore$ is the core mass, $\Rcore$ is the core radius, $G$ is the gravitational constant, and $\delta \vturb$ is the turbulent velocity. The core mass is estimated from the JCMT $850~\micron$ flux, which is presented by \cite{Yen:inprep}, with the following equation:
\begin{equation}
    M_\mathrm{850} = \frac{F_\nu d^2}{\kappa_\nu B_\nu (T_\mathrm{dust})},
\end{equation}
where $F_\nu$ is the flux density, $d$ is the distance, $\kappa_\nu$ is the dust opacity, $B_\nu$ is the Planck function, and $T_\mathrm{dust}$ is the dust temperature. $F_\nu$ is measured to be $2.9~\Jy$ with a typical flux uncertainty of 10\%. We adopt the dust opacity $\kappa_\nu = 0.02~\mathrm{g^{-1} cm^2}$ \citep{Ossenkopf:1994aa}, and the dust temperature of $15\pm5~\kelv$. The core radius is estimated to be $6700\pm800~\au$ from the $3\sigma$ contour of the $850~\micron$ continuum emission \citep{Yen:inprep}. The uncertainty of the core radius is the standard deviation of the radii of the $3\sigma$ contour, which vary with the azimuthal angle. We used the $2'$ square map of the C$^{18}$O $J=2$--1 emission presented by \cite{Sai:2023aa} to measure the turbulent velocity in two ways. One way is to measure the full width at half maximum (FWHM) of a spectrum and subtract the thermal line width. The FWHM of the spectrum is measured to be $0.50\pm0.02~\kmps$ within a circle with a radius of $8''$ at an off position, $40''$ north of the protostellar position, where systemic motions such as infall or rotation are not observed. By assuming the gas temperature same as the dust temperature above, the non-thermal velocity ($\sigma_\mathrm{nth}$) is estimated to be $0.20 \pm 0.02~\kmps$. This would give an upper limit of the turbulent velocity of the dense core, as the C$^{18}$O $J=2$--1 emission may include the surrounding cloud component. These parameters yield $\beturb \tsim 1.4\pm0.9$, where the uncertainty is calculated through the error propagation of the uncertainty of each parameter. The calculated turbulent energy against the gravitational energy is much larger than those in numerical simulations reporting the time variation of the orientation of the disk normal.

The above calculation of the turbulent velocity may overestimate the turbulence energy in the dense core. The other way to estimate the turbulent velocity is to calculate the velocity deviation in the moment 1 map across the dense core diameter, since it is suggested that the velocity gradient on scales of $\gtrsim 1000~\au$ originates from turbulent motion \citep{Sai:2023aa}. \cite{Sai:2023aa} provide a scaling relation between the velocity deviation ($\delta v$) and the spatial scale ($\tau$) of
\begin{equation}
    \delta v = 10^{-1.69\pm0.08} \left( \frac{\tau}{1000~\au} \right)^{0.6\pm0.2} ~ (\kmps).
    \label{eq:delv_15398}
\end{equation}
This relation could provide a better estimation on the turbulent velocity on a specific spatial scale, because the line-of-sight components that could be larger than the map area are averaged out in the moment 1 map. By adopting this scaling relation and $\tau=2\Rcore$, the turbulent velocity across the dense core is estimated to be $0.10 \pm 0.01~\kmps$. This estimate results in $\beturb \tsim 0.33 \pm 0.20$, which is also larger than those of the simulations. Hence, considering the strength of the turbulence in the dense core, turbulence could cause the dynamical time variation of the outflow direction in the dense core of IRAS 15398.


\subsubsection{Misalignment Scenario} \label{subsec:msal_scenario}
A misalignment between the rotational axis of the initial dense core and the associated magnetic field could result in a significant change in the outflow direction over time. \cite{Machida:2020aa} show that the orientations of the disk normal and the low-velocity outflow ($<10~\kmps$) change by $\rtsim90^\circ$ when the initial misalignment angle between the magnetic field and the rotational axis of the dense core is $> 80^\circ$.

\begin{figure}
\centering
\plotone{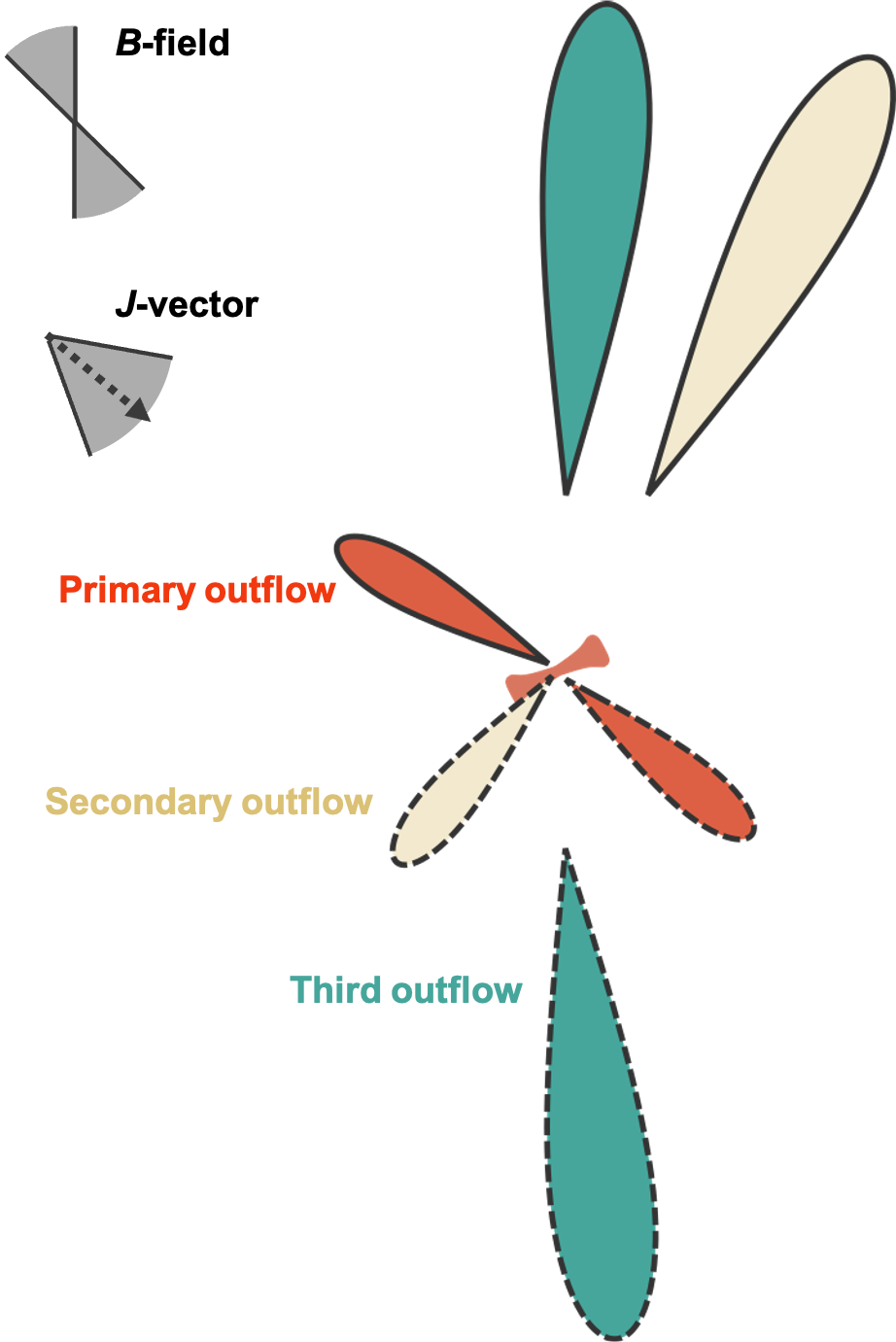}
\caption{Summary of the configurations of the multiple outflows and the possible orientations of the magnetic field and the angular momentum vector of the dense core on the plane of the sky. Lobes drawn with solid and dashed curves indicate the redshifted and blueshifted components of the outflows, respectively. \label{fig:summary_config}}
\end{figure}

Previous observations with SOFIA reveal orientations of magnetic fields on scales of $\rtsim9000~\au$ around IRAS 15398, providing a mean position angle of the magnetic field is $\rtsim225^\circ$ \citep[][]{Redaelli:2019aa}. On the other hand, the JCMT/POL-2 map of $850~\micron$ continuum emission suggests that the overall direction of the magnetic field on scales of $\rtsim6000~\au$ is north-south \citep[P.A. of $180^\circ$; ][]{Yen:inprep}. The angular momentum vector of the initial dense core could be estimated from the velocity gradient on scales of $\rtsim6000$--$9000~\au$ around IRAS 15398. \cite{Sai:2023aa} have revealed that the P.A. of the velocity gradient varies from $\rtsim107$ to $166^\circ$ at radii of $6000$--$9000~\au$, using the C$^{18}$O $J=2$--1 emission. These gradient directions imply that P.A. of the angular momentum vector of the dense core is $\rtsim200$--$260^\circ$, which is orthogonal to the gradient directions. In Figure \ref{fig:summary_config}, we summarize these possible orientations of the magnetic field and the angular momentum vector of the dense core on the plane of the sky with the configurations of the multiple outflows.

The misalignment angle between the magnetic field and the angular momentum vector can be 20--$35^\circ$ on the plane of the sky, when comparing the magnetic field orientation and velocity gradient measured on the same spatial scale. The probability of a significant misalignment between the magnetic field and the rotational axis of the dense core in 3D space, given a projected misalignment angle, can be estimated through a statistical approach. Assuming that the magnetic field and the rotational axis of the dense core are randomly oriented in 3D space, the possibility that the misalignment angle is $>80^\circ$ in 3D space and appears to be 20--$35^\circ$ on the plane of the sky after projection is less than 10\% \citep{Galametz:2018aa, Gupta:2022aa}. Thus, the magnetic field is less likely largely misaligned with the rotational axis of the dense core of IRAS 15398, although the possibility of a large misalignment of $>80^\circ$ cannot be completely ruled out.

\subsection{How Common are Multiple Outflows around Single Protostars?}
\subsubsection{Is IRAS 15398 a Peculiar Source?}
Multiple, highly-misaligned outflows possibly launched by single protostars, such as those found in IRAS 15398, have not been observed except for very few recent reports \citep{Okoda:2021aa, Sato:2023aa}, although it should be noted that multiple outflows have not been searched for in a systemic way. As discussed in the previous section, the multiple outflows in IRAS 15398 could be attributed to turbulence in its dense core. A question worth discussing is whether the turbulence in the dense core of IRAS 15398 is notably stronger compared to that in other dense cores.

\begin{figure*}
\epsscale{1.1}
\centering
\plotone{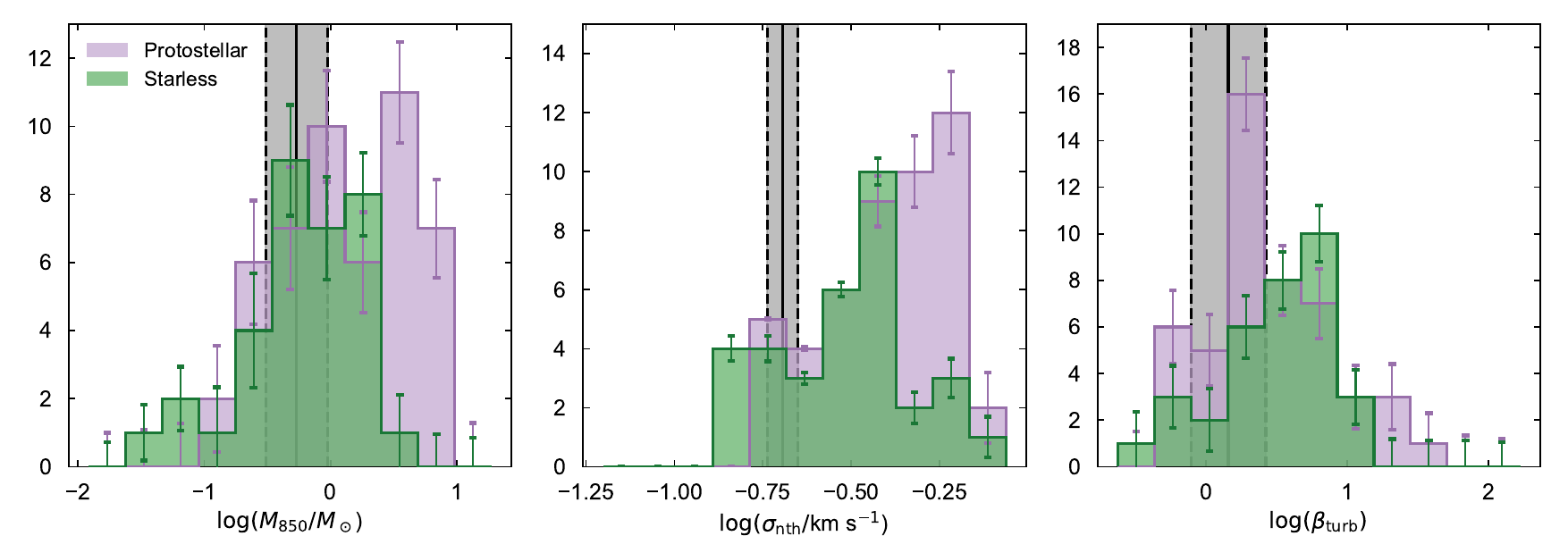}
\caption{Histograms of the mass, non-thermal velocity, and the ratio of the turbulence energy to the gravitational energy of the dense cores in the Perseus star forming region from left to right, which are referred from \cite{Kirk:2007aa}. Vertical lines show the parameters for the dense core of IRAS 15398 and grey shaded area indicate their uncertainties. Non-thermal velocities are measured with the FWHM of the $\CeighteenO$ $J=2$--1 spectrum for both Perseus dense cores and the dense core of IRAS 15398.\label{fig:hist-coreprop}}
\end{figure*}

We compare the turbulent velocity and energy of the dense core of IRAS 15398 with those of the dense cores in the Perseus star-forming regions, which have been well characterized with comprehensive surveys in both (sub)mm line and continuum emission \citep{Kirk:2006aa, Kirk:2007aa}. To calculate the turbulent energy and the gravitational energy for the Perseus dense cores, we compiled the JCMT $850~\micron$ fluxes, the core radii and the total line widths of the $\CeighteenO$ $J=2$--1 and $\NtwoHp$ $J=1$--0 lines from the catalog provided by \cite{Kirk:2007aa}. Then, we calculated the core masses and non-thermal velocities assuming the same dust opacity and temperature as for IRAS 15398. We adopted the distance of $282 \pm28~\pc$ for the Perseus dense cores, derived from the mean and standard deviation of Gaia distances to different regions of the Perseus star-froming region \citep{Zucker:2022aa}. Note that \cite{Kirk:2007aa} identified the dense cores using the \texttt{Clumpfind} algorithm and adopted the $3\sigma$ contour edge as the definition of the core radius. Our definition of the core radius for IRAS 15398 is almost identical with theirs, and the rms of the $850~\micron$ map for IRAS 15398 falls within a typical range of rms noises of their $850~\micron$ maps. First, the $\CeighteenO$ non-thermal velocities ($\sigma_\mathrm{nth}$) and $\beturb$ calculated with them are compared between IRAS 15398 and the Perseus dense cores in Figure \ref{fig:hist-coreprop}. The core mass, turbulent velocity and $\beturb$ of IRAS 15398 are comparable to or less than the typical values in both the starless and protostellar dense cores in the Perseus star-forming region.

\begin{figure*}
\epsscale{1.1}
\centering
\plotone{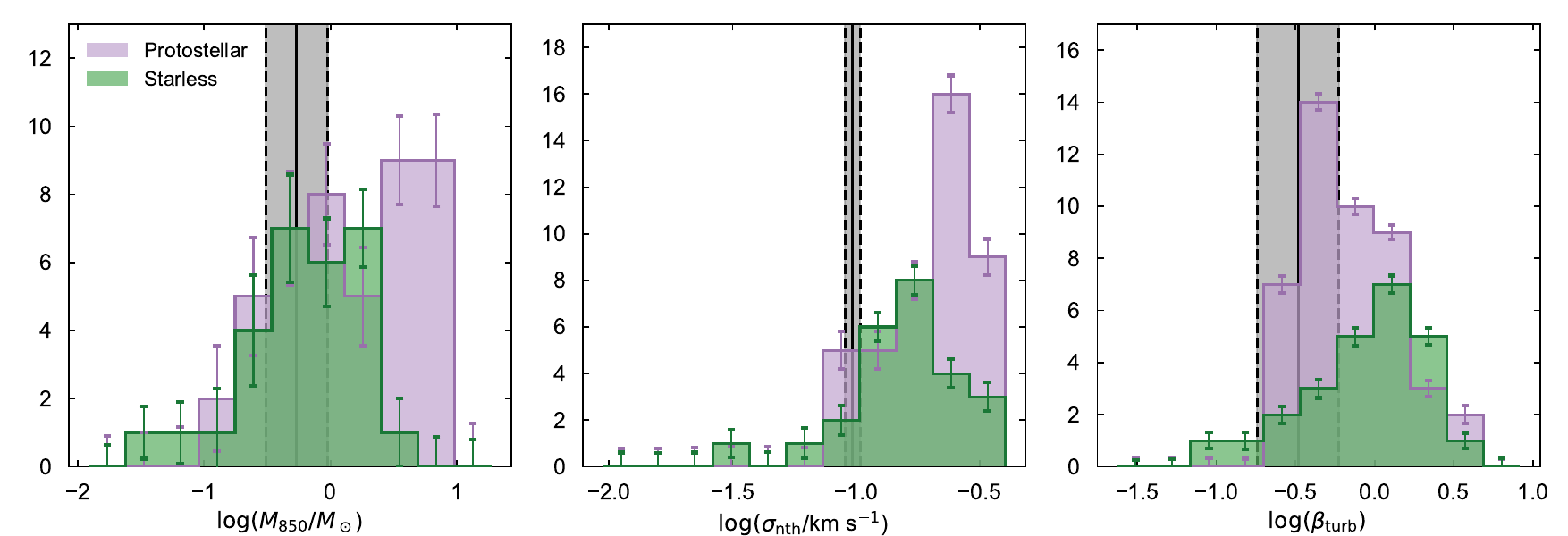}
\caption{Same as Figure \ref{fig:hist-coreprop} but with measurements using the $\NtwoHp$ $J=$1--0 emission for the dense cores in the Perseus star-forming region, referred from \cite{Kirk:2007aa}, and with the non-thermal velocity measured using the scaling relation of Equation (\ref{eq:delv_15398}) for IRAS 15398. \label{fig:hist-coreprop-n2hp}}
\end{figure*}

The $\CeighteenO$ emission observed by single-dish telescopes may include the surrounding cloud components, and the turbulent velocity could be overestimated. Thus, we performed similar comparisons using the turbulent velocities of the Perseus dense cores measured with the $\NtwoHp$ emission, which typically traces denser regions than the $\CeighteenO$ emission \citep{Kirk:2007aa}. Because there is no available $\NtwoHp$ data taken by single-dish telescopes for IRAS 15398, we used the turbulent velocity measured with the scaling relation of Equation (\ref{eq:delv_15398}). Line widths of the $\NtwoHp$ emission of dense cores are typically smaller than those of the $\CeighteenO$ emission by a factor of two \citep{Johnstone:2010aa}. The turbulent velocity measured with Equation (\ref{eq:delv_15398}) is also about twice smaller than that measured with the $\CeighteenO$ emission, suggesting that our method is tracing similar spatial scales to those of the $\NtwoHp$ emission. 
Figure \ref{fig:hist-coreprop-n2hp} presents histograms of these parameters. The core mass, turbulent velocity, and $\beturb$ of IRAS 15398 are similar to or less than the typical values in the Perseus dense cores. These comparisons suggest that turbulence of the dense core of IRAS 15398 is not notably stronger compared to other dense cores. Furthermore, most of the protostellar and starless cores in the Perseus star-forming region exhibit $\beturb \gtrsim 0.1$, which is larger than those of numerical simulations reporting time evolution of the disk and/or outflow orientations \citep{Seifried:2013aa, Matsumoto:2017aa}. This may suggest that protostellar systems in these dense cores also experience changes in disk and outflow directions over time, which contradicts the few reports of multiple outflows.


\subsubsection{Age of the System and Outflow Dissipation Timescale}
Based on the results of the numerical simulations and the observed distributions of the turbulence energy of the dense cores, we would expect that the disk/outflow orientations should evolve with time and multiple outflows form in many sources. A possible explanation for few reports of multiple outflows is that the significant change in the outflow direction only occurs within a short period at an early phase of the star formation process. While the disk would be small and its orientation would be easily altered at the beginning of the star formation process, it could be difficult to change the orientation of a larger, more massive disk at later evolutionary stages. If this is the case and the outflow launched along different directions in an early phase also dissipates within a similar timescale, then multiple outflows would be only observed in young sources.

IRAS 15398 has been considered to be at an early stage of the protostellar phase because of its small protostellar mass \citep[][]{Okoda:2018aa, Thieme:2023aa}. \cite{Thieme:2023aa} estimated the protostellar age ($t_\ast$) from the protostellar mass and mass accretion rate ($\Macc$) by $t_\ast = M_\ast / \Macc$, assuming a constant accretion rate over time. They obtained the mass accretion rate of $(1.3$--$6.1)\times 10^{-6}~\msunpyr$ adopting the accretion luminosity of $\Lacc \tsim \lbol = 1.4~\Lsun$, $M_\ast\tsim0.02$--$0.1~\Msun$, and the protostellar radius of $R_\ast = 3~\Rsun$ through the equation of $\Macc = \Lacc R_\ast/ (G M_\ast)$, where $G$ is the gravitational constant.
This mass accretion rate and the protostellar mass yield $t_\ast \rtsim (0.4$--$7.5)\times10^{4}~\yr$. The actual mass accretion rate would be variable over time. Theoretical works predict that the mass accretion rate for low-mass protostars is typically a few $\times10^{-6}$--$10^{-5}~\msunpyr$ and beyond $10^{-5}~\msunpyr$ when strong accretion outbursts occur via the disk gravitational instability \citep{Dunham:2012aa}. The mass accretion rate estimated for IRAS 15398 with the current luminosity is closer to the accretion rate at a steady state in the numerical simulations. The actual protostellar age could be further smaller than the above estimate, if the mass accretion rate of IRAS 15398 is higher during some periods due to accretion outbursts. \cite{Thieme:2023aa} also estimated the protostellar age to be $t_\ast \rtsim (0.6$--$1.9)\times10^{4}~\yr$ from comparison between the specific angular momentum measured for the envelope of IRAS 15398 and a core collapse model provided by \cite{Takahashi:2016aa}, which assumes the angular momentum conservation during the core collapse. These protostellar ages of $\rtsim 10^{4}~\yr$ are smaller than the typical lifetime of the Class 0 protostars of $\rtsim10^5~\yr$ \citep{Kenyon:1990aa, Evans:2009aa, Dunham:2015aa}, suggesting that IRAS 15398 is at an early stage of the protostellar phase.

The dynamical timescales of the secondary and third outflows are $\tau_\mathrm{dyn} \gtrsim 0.5$--$1\times 10 ^4~\yr$, which is comparable to the protostellar age of IRAS 15398. This suggests that the second and third outflows have been ejected at a very early phase just after formation of the protostar. The misaligned outflows ejected at such an early phase would dissipate within a short timescale, because there is no more momentum injection in the same direction. The dissipation timescale of the observed outflows can be estimated by equating their momentum and the drag force that the outflows experience \citep{Codella:1999aa}. The drag force can be written as $F_\mathrm{d} = \frac{1}{2} C_D \rho_\mathrm{core} v_\mathrm{out}^2 S$, where $C_D$ is a drag coefficient, $\rho_\mathrm{core}$ is the density of the surrounding dense core, $v_\mathrm{out}$ is the outflow velocity and $S$ is the cross section area. The momentum of the outflow is $P_\mathrm{out} = M_\mathrm{out} v_\mathrm{out}$, where $P_\mathrm{out}$ and $M_\mathrm{out}$ are the momentum and mass of the outflow, respectively. By relating these two equations, the dissipation timescale of the outflow is obtained as:
\begin{eqnarray}
    t_\mathrm{diss} = \frac{P_\mathrm{out}}{F_D} & \rtsim 2 \frac{\rho_\mathrm{out}}{\rho_\mathrm{core}} \frac{l_\mathrm{out}}{v_\mathrm{out}} \\
    & = 2 \frac{\rho_\mathrm{out}}{\rho_\mathrm{core}} \frac{l_\mathrm{out}'}{v_\mathrm{out}'} \frac{1}{\tan i},
\end{eqnarray}
where $\rho_\mathrm{out}$ is the mean density of the outflow, $l_\mathrm{out}$ is the outflow length in 3D space, $l_\mathrm{out}'$ and $v_\mathrm{out}'$ are the outflow length and velocity on the plane of the sky, and $i$ is the inclination angle.  Here, we assume $C_D=1$ \citep{Codella:1999aa} and $M_\mathrm{out} \tsim \rho_\mathrm{out} l_\mathrm{out} S$. 
The typical core density is $n_\mathrm{core} \tsim 1.0 \times 10^4~\perccm$ \citep[e.g.,][]{Ward-Thompson:2007ab}, corresponding to $\rho_\mathrm{core} \tsim 4.7\times 10^{-20}~\gperccm$ assuming the mean molecular weight of 2.8 \citep{Kauffmann:2008aa}. An averaged mass of the outflows are estimated to be $\rtsim(0.32\mbox{--}1.1)\times10^{-3}~\Msun$ from the mean integrated flux of $\rtsim96~\Jy~\kmps$ (Figure \ref{fig:maps_diffvel}), assuming $\tex=50\mbox{--}200~\kelv$ \citep{van-Kempen:2009ac, Yildiz:2012aa, Yildiz:2013aa, Dunham:2014aa}, a CO molecular abundance of $10^{-4}$, the mean molecular weight of 2.8 and the local thermal equilibrium (LTE). Note that the mass of the outflows is likely underestimated, because we do not take into account low-velocity components obscured by the optically thick, extended foreground gas. \cite{Offner:2011aa} found that the outflow mass is underestimated by a factor of $\rtsim5$--$10$ by neglecting low-velocity emission with $|v| \lesssim 2~\kmps$ based on synthetic observations of numerical simulations. Hence, we adopt an outflow mass of $\rtsim 1 \times10^{-2}~\Msun$ for an order of magnitude estimate. By adopting outflow lengths of $\rtsim8500~\au$ (or $\rtsim55''$) and widths of $\rtsim1600~\au$ (or $\rtsim10''$) based on Figure \ref{fig:maps_diffvel}, the outflow density is calculated to be $\rtsim9 \times 10^{-20}~\gperccm$. The ratio of the outflow density to the core density ($\rho_\mathrm{out}/\rho_\mathrm{core}$) of $\rtsim 2$ calculated from the above densities is roughly consistent with the density ratio found in numerical simulations \citep{Machida:2012aa}. The density ratio, the outflow length, the outflow velocity of $\rtsim3~\kmps$ and the inclination angle of $i\tsim70^\circ$ yield the dissipation timescale of $\rtsim 2 \times 10^4~\yr$. The calculated dissipation timescale is comparable with the dynamical timescale and the age of the system. Thus, these misaligned, multiple outflows that may form at an early stage of the star formation process would dissipate on a similar timescale to the current age of IRAS 15398.

As discussed above, dense cores including IRAS 15398 commonly exhibit turbulent energy large enough to form multiple outflows, and a possible reason why the multiple outflows are rarely observed is that they form only at an early stage of the star formation process and dissipate in a short timescale.
The ratio of the age of IRAS 15398 to the typical lifetime of protostars is about $\rtsim1/10$, suggesting that one of ten protostellar systems may have similar multiple outflows. Systemic investigations on presence/absence of multiple outflows around single protostars are required to further test this hypothesis.

\section{Conclusions} \label{sec:summary}
We reported our finding of the third outflow, as well as the possible redshifted lobe of the secondary outflow, around a single protostar IRAS15398$-$3359 based on mosaic observations in the CO $J=2$--1 emission using the ACA 7-m array. Main results and conclusions are summarized below:
\begin{enumerate}
    \item Our ACA 7-m array observations have newly revealed low-velocity lobes or filamentary structures at distances from the protostar larger than $\rtsim30''$ in addition to the known high-velocity primary and blueshifted secondary outflows.
    \item Two lobes, named as NR1 and SB, appear to be symmetric with respect to the protostar and exhibit Hubble-law like velocity profiles, suggesting that the NR1 and SB lobes likely trace the third outflow. The SEB lobe corresponds to the known secondary outflow. The NR2 lobe, which aligned in a line passing through the SEB lobe and the protostar, also exhibits a Hubble-law like velocity profile, suggesting that this is the counterpart of the blueshifted secondary outflow. 
    \item The dynamical timescales of the third outflow is estimated to be $\gtrsim 5 \times 10^3~\yr$. It suggests that the third outflow would be the oldest one and followed by the primary outflows with the dynamical timescale of $\rtsim500~\yr$. These multiple outflows are misaligned with each other by $\rtsim20$--$90^\circ$ on the plane of the sky. IRAS 15398 has been found to be a single protostar with the observation at $\rtsim6~\au$ resolution. These observations suggest that the highly-misaligned, multiple outflows are a result by the evolution of the disk and outflow orientations.
    \item The ratio of the turbulent energy to the gravitational energy in the dense core of IRAS 15398 is calculated to be $\beta\tsim 0.33$--$1.4$ based on previous observations. The turbulent energy in IRAS 15398 is sufficiently large compared to those in numerical simulations reporting the evolution of the disk/outflow orientations, while the magnetic field is unlikely to be largely misaligned ($\gtrsim80^\circ$) with the rotational axis of the dense core. Hence, the multiple outflows in IRAS 15398 could be attributed to the turbulence in its dense core.
    \item The turbulent energy of the dense core of IRAS 15398 is within a typical range of those of other dense cores in the Perseus star-forming region. These comparisons hint that the significant change in the disk and outflow directions over time can be a common event at the early stage of the star formation process. The few identifications of multiple outflows could be due to the short timescale of the formation and dissipation of the misaligned outflows.
\end{enumerate}

\begin{acknowledgments}
This paper used the following ALMA data: ADS/JAO.ALMA \#2019.1.01063.S. ALMA is a partnership of ESO (representing its member states), NSF (USA), and NINS (Japan), together with NRC (Canada), MOST and ASIAA (Taiwan), and KASI (Republic of Korea), in cooperation with the Republic of Chile. The Joint ALMA Observatory is operated by ESO, AUI/NRAO, and NAOJ. We thank all ALMA staff for conducting the observations. We also thank Yuki Okoda for the discussion on the interpretation of the observational data. H.-W.Y.~acknowledges support from the National Science and Technology Council (NSTC) in Taiwan through the grant NSTC 1102628-M-001-003-MY3 and from the Academia Sinica Career Development Award (AS-CDA-111-M03). N.O.~acknowledges support from the NSTC through the grant NSTC 109-2112-M-001-051, 110-2112-M-001-031.
\end{acknowledgments}

%

\vspace{5mm}
\facilities{ALMA}


\software{CASA \citep{McMullin:2007aa}, Numpy \citep{Oliphant:2006aa,van-der-Walt:2011aa}, Scipy \citep{Virtanen:2020aa}, Astropy \citep{Astropy-Collaboration:2013aa,Astropy-Collaboration:2018aa}, Matplotlib \citep{Hunter:2007aa}, \texttt{emcee} \citep[][]{Foreman-Mackey:2013aa}}



\restartappendixnumbering
\appendix

\section{Supplemental Figures}
\begin{figure*}
\epsscale{1.2}
\plotone{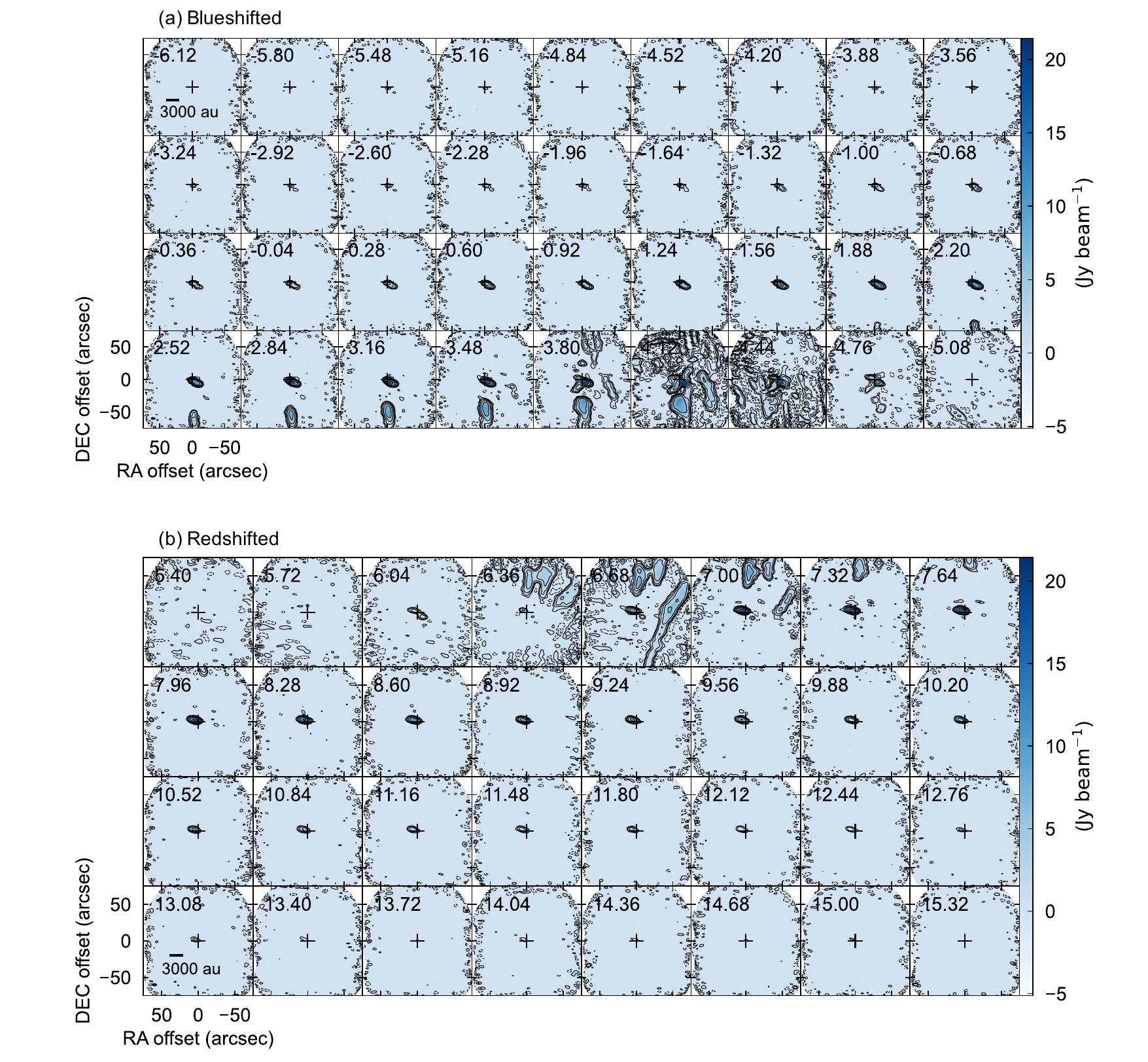}
\caption{Velocity channel maps of (a) blueshifted and (b) redshifted components of the CO $J=2$--1 emission. Contour levels are 3, 6, 12, 24, ... $\times\sigma$, where 1$\sigma$ is $0.13~\jypbm$. Maps are shown in steps of two channels. The label at the top left corner in each panel denotes the LSR velocity of the channel in $\kmps$. The central crosses denote the protostellar position. The ellipse at the bottom left corner indicates the synthesized beam size of $7\farcs7\times 4\farcs1$ (87$^\circ$). \label{fig:channels}}
\end{figure*}

The velocity channel maps of the CO $J=2$--1 emission are presented in Figure \ref{fig:channels}.

\section{Wind-driven Shell Model Fitting} \label{app:modelfit}
\begin{figure*}
\epsscale{1.2}
\plotone{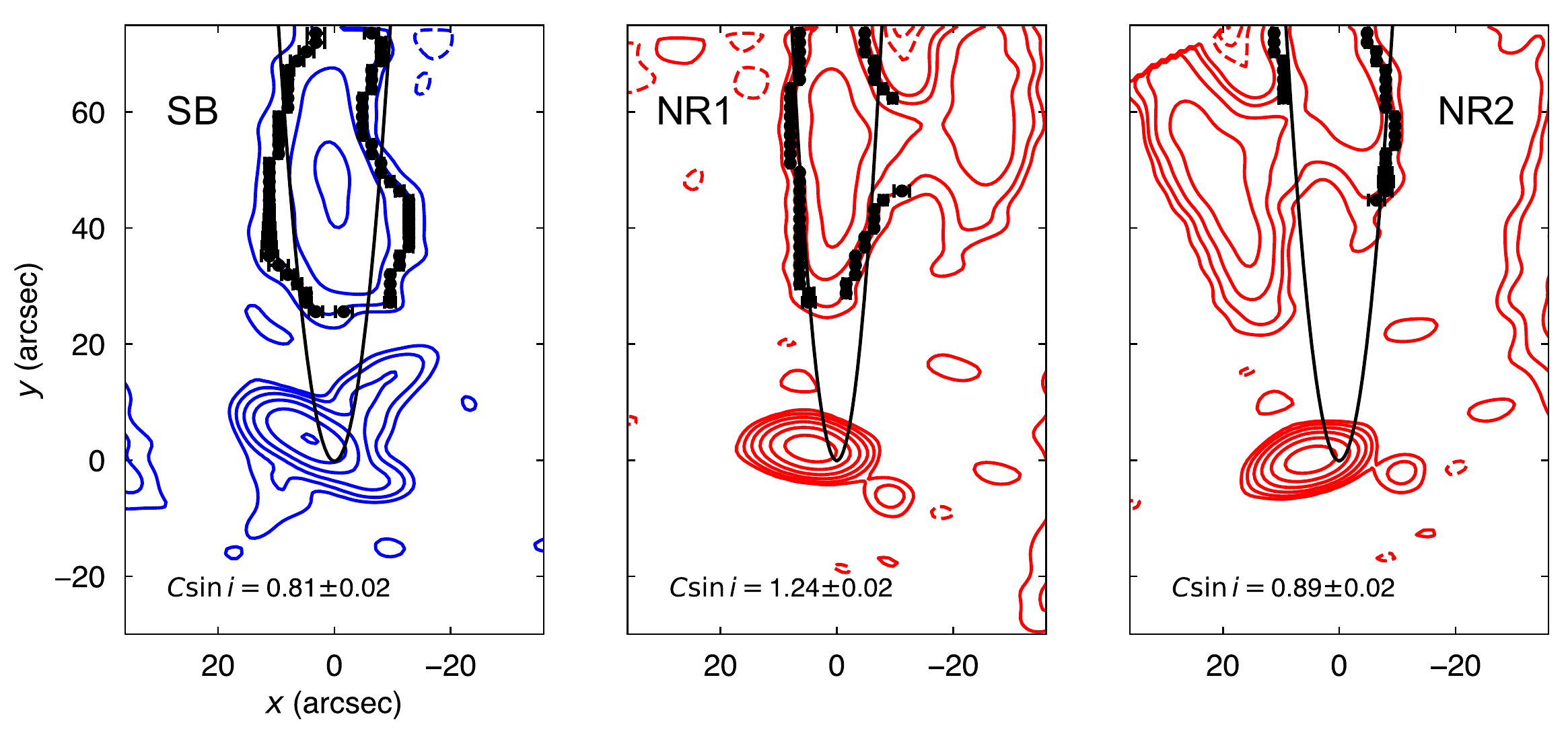}
\caption{Results of the fitting of the outflow shell morphology. Black points are measured data points corresponding to $5\sigma$ contours. Black curves indicate the best-fit outflow shell morphologies.\label{fig:app_cavityfit}}
\end{figure*}

\begin{figure*}
\plotone{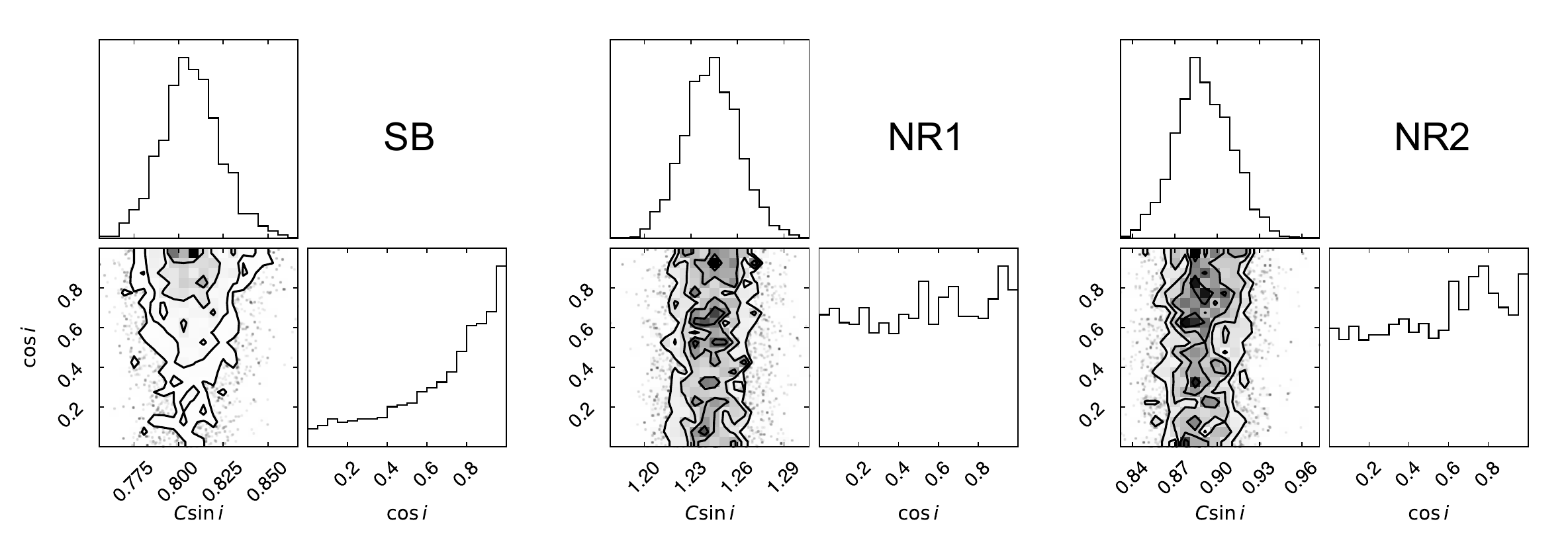}
\caption{Posterior distributions of the parameters for the fitting of the shell morphology. \label{fig:app_cavityfit_corners}}
\end{figure*}

We present the fitting method of the wind-driven shell model \citep{Shu:1991aa, Lee:2000aa} in this section. The method consists of two steps: fitting of the morphology and fitting of the velocity structure. In the first step, we fit the outflow shell morphology using the moment 0 maps presented in Figure \ref{fig:maps_diffvel}. We use the $5\sigma$ contours of the observed emission to trace outflow shell morphologies. First, the moment 0 maps are rotated by P.A. of the outflow lobes, which is measured as lines passing through the protostellar position and the emission peak of the outflow lobes, so that the outflow axis corresponds to the $y$ axis of the plane of the sky (Figure \ref{fig:app_cavityfit}). Then, we extract intensity profiles along the $x$ axis at $y$ offsets sampled by a half of the beam size, and measure positions of the intensity at the $5\sigma$ level. Uncertainties of the measured positions are calculated as positional offsets at $5\pm1\sigma$ contours from the $5\sigma$ contour. The NR1 and NR2 lobes spatially overlap. Therefore, we separate them at local intensity minima of the extracted intensity profiles. At a given $y$ offset, two $x$ values corresponding to the outflow shell edges on different sides are obtained. To better trace the outflow shell morphology, the inner data point is removed when both two points are on the same side (i.e., when both are $x<0$ or $x>0$). Similarly, data points at artificial edges, i.e., the local intensity minima where the lobes are separated, are removed. 
The projected outflow shell morphology is written as
\begin{eqnarray}
    y = C \sin i x^2 - \frac{\cos^2 i}{4C \sin i}, \nonumber \\
    \therefore x = \pm \sqrt{\frac{1}{C \sin i} y + \frac{\cos^2 i}{4 C^2 \sin^2 i}}, \label{eq:_csini}
\end{eqnarray}
where $x$ and $y$ are plane-of-sky coordinates. The inclination angle is defined as an angle between the line-of-sight axis and the outflow axis. We search for the parameter set of $(C\sin i, \cos i)$ to best explain the observations by fitting the data points with the Markov-Chain Monte Carlo (MCMC) code \texttt{emcee} \citep{Foreman-Mackey:2013aa}. Uncertainties of the fitting parameters are calculated as the standard deviation of the posterior distributions. The obtained data points and fitting results are shown in Figure \ref{fig:app_cavityfit}. The posterior distributions are also presented in Figure \ref{fig:app_cavityfit_corners}. The cavity shape, $C \sin i$, is well constrained, while the cosine of the inclination angle, $\cos i$, does not converge. Thus, we leave the inclination angle as a free parameter in the second fitting step.

\begin{figure*}
\plotone{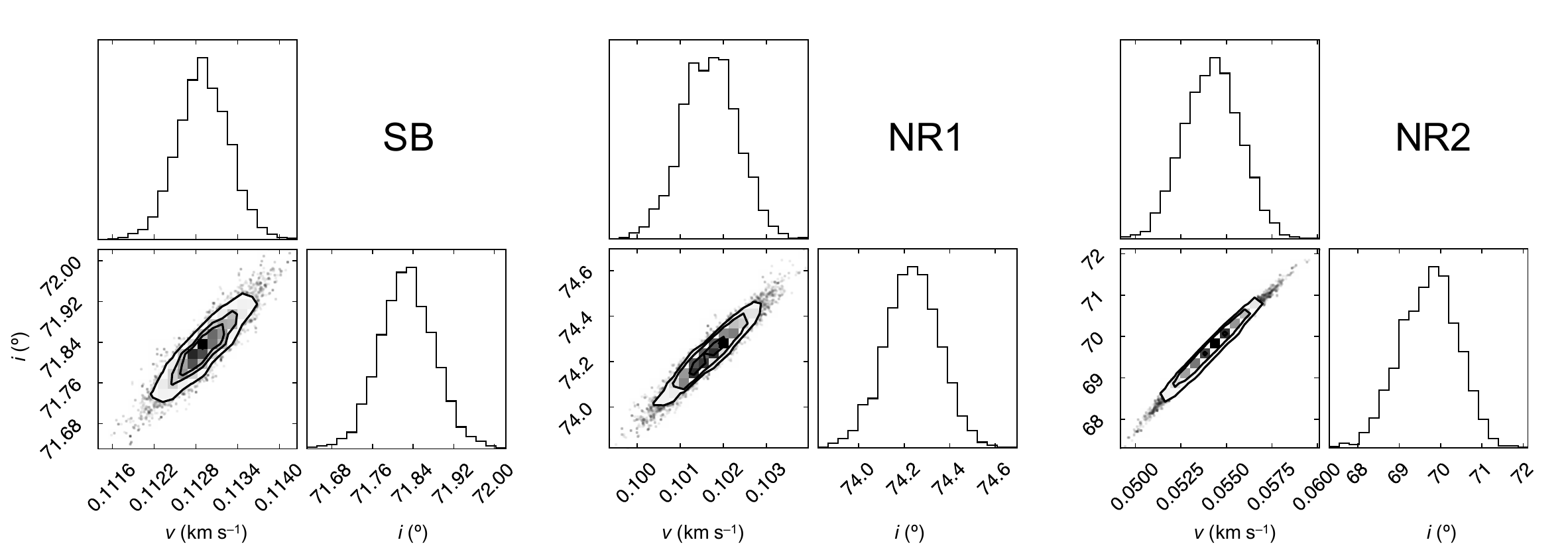}
\caption{Same as Figure \ref{fig:app_cavityfit_corners} but for the fitting of the velocity structure.\label{fig:app_pvfit_corners}}
\end{figure*}

\begin{figure*}
\plotone{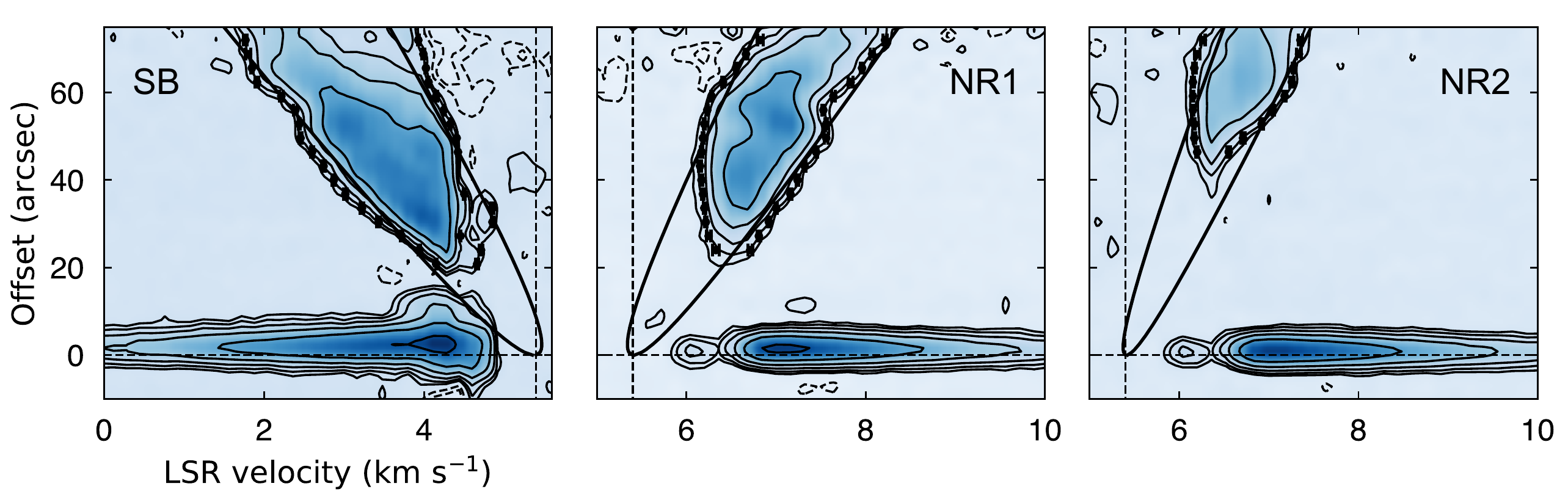}
\caption{Same as Figures \ref{fig:pvds-3rdoutflow} and \ref{fig:pvd-nr2} but with fitting results for all data points including those likely affected by the cloud contamination. \label{fig:app_pvfit}}
\end{figure*}

In the second step, we fit the velocity structure using PV diagrams and $C \sin i$ constrained in the first step. The $5 \pm 1\sigma$ contours are used to measure velocities and their uncertainties at offsets sampled by a half of the beam size. Around the systemic velocity, the emission is highly affected by the cloud contamination and its resolve-out. Thus, data points at velocities less than $4.4~\kmps$ are removed from the fitting for the SB lobe to better trace the velocity structure of the outflow. Similarly, data points at velocities larger than $6.34~\kmps$ and $6.25~\kmps$ are removed from fits for the NR1 and NR2 lobes, respectively. 
We perform the fitting of the velocity structure in the same manner as in the first step. The posterior distributions of the parameters are shown in Figure \ref{fig:app_pvfit_corners}, and fitting results are presented in Figures \ref{fig:pvds-3rdoutflow} and \ref{fig:pvd-nr2}, and in Table \ref{tab:res-modelfit}. Note that these fitting results do not change significantly even if we include the data points that are likely affected by the cloud contamination. The fitting results using all the data points are presented in Figure \ref{fig:app_pvfit} and Table \ref{tab:app-fitpvres}. The uncertainties of $v_0$ and $i$ shown in Table \ref{tab:res-modelfit} and \ref{tab:app-fitpvres} do not take into account systemic errors of the fitting method itself. Based on the differences of $v_0$ and $i$ values between Table \ref{tab:res-modelfit} and \ref{tab:app-fitpvres}, systemic errors of $v_0$ and $i$ associated with the selection of the data point around the cloud velocity are about 10\% and 1\%, respectively.

\begin{deluxetable}{lccc}
\tablecaption{Results of Wind-driven Shell Model Fits for All Data Points} \label{tab:app-fitpvres}
\tablehead{
\colhead{Lobes} & \colhead{$C \sin{i}$} & \colhead{$v_0$} & \colhead{$i$} \\
\colhead{} & \colhead{} & \colhead{($\kmps$)} & \colhead{($^\circ$)}}
\startdata
SB  & $0.81\pm0.02$ & $0.1007\pm0.0003$ & $69.35\pm0.05$ \\
NR1 & $1.24 \pm 0.02$ & $0.0963\pm0.0004$ & $73.22\pm0.06$ \\
NR2 & $0.89\pm 0.02$ & $0.0487\pm0.0005$ & $66.66\pm0.20$
\enddata
\end{deluxetable}

\clearpage
\bibliography{reference,reference_spl}{}

\begin{thebibliography}{}
\expandafter\ifx\csname natexlab\endcsname\relax\def\natexlab#1{#1}\fi
\providecommand{\url}[1]{\href{#1}{#1}}
\providecommand{\dodoi}[1]{doi:~\href{http://doi.org/#1}{\nolinkurl{#1}}}
\providecommand{\doeprint}[1]{\href{http://ascl.net/#1}{\nolinkurl{http://ascl.net/#1}}}
\providecommand{\doarXiv}[1]{\href{https://arxiv.org/abs/#1}{\nolinkurl{https://arxiv.org/abs/#1}}}

\bibitem[{{Allen} {et~al.}(2003){Allen}, {Li}, \& {Shu}}]{Allen:2003aa}
{Allen}, A., {Li}, Z.-Y., \& {Shu}, F.~H. 2003, \apj, 599, 363,
  \dodoi{10.1086/379243}

\bibitem[{{Alves} {et~al.}(2017){Alves}, {Girart}, {Caselli}, {Franco}, {Zhao},
  {Vlemmings}, {Evans}, \& {Ricci}}]{Alves:2017aa}
{Alves}, F.~O., {Girart}, J.~M., {Caselli}, P., {et~al.} 2017, \aap, 603, L3,
  \dodoi{10.1051/0004-6361/201731077}

\bibitem[{{Andrews}(2020)}]{Andrews:2020aa}
{Andrews}, S.~M. 2020, \araa, 58, 483,
  \dodoi{10.1146/annurev-astro-031220-010302}

\bibitem[{{Astropy Collaboration} {et~al.}(2013){Astropy Collaboration},
  {Robitaille}, {Tollerud}, {Greenfield}, {Droettboom}, {Bray}, {Aldcroft},
  {Davis}, {Ginsburg}, {Price-Whelan}, {Kerzendorf}, {Conley}, {Crighton},
  {Barbary}, {Muna}, {Ferguson}, {Grollier}, {Parikh}, {Nair}, {Unther},
  {Deil}, {Woillez}, {Conseil}, {Kramer}, {Turner}, {Singer}, {Fox}, {Weaver},
  {Zabalza}, {Edwards}, {Azalee Bostroem}, {Burke}, {Casey}, {Crawford},
  {Dencheva}, {Ely}, {Jenness}, {Labrie}, {Lim}, {Pierfederici}, {Pontzen},
  {Ptak}, {Refsdal}, {Servillat}, \&
  {Streicher}}]{Astropy-Collaboration:2013aa}
{Astropy Collaboration}, {Robitaille}, T.~P., {Tollerud}, E.~J., {et~al.} 2013,
  \aap, 558, A33, \dodoi{10.1051/0004-6361/201322068}

\bibitem[{{Astropy Collaboration} {et~al.}(2018){Astropy Collaboration},
  {Price-Whelan}, {Sip{\H o}cz}, {G{\"u}nther}, {Lim}, {Crawford}, {Conseil},
  {Shupe}, {Craig}, {Dencheva}, {Ginsburg}, {VanderPlas}, {Bradley},
  {P{\'e}rez-Su{\'a}rez}, {de Val-Borro}, {Aldcroft}, {Cruz}, {Robitaille},
  {Tollerud}, {Ardelean}, {Babej}, {Bach}, {Bachetti}, {Bakanov}, {Bamford},
  {Barentsen}, {Barmby}, {Baumbach}, {Berry}, {Biscani}, {Boquien}, {Bostroem},
  {Bouma}, {Brammer}, {Bray}, {Breytenbach}, {Buddelmeijer}, {Burke},
  {Calderone}, {Cano Rodr{\'{\i}}guez}, {Cara}, {Cardoso}, {Cheedella},
  {Copin}, {Corrales}, {Crichton}, {D'Avella}, {Deil}, {Depagne}, {Dietrich},
  {Donath}, {Droettboom}, {Earl}, {Erben}, {Fabbro}, {Ferreira}, {Finethy},
  {Fox}, {Garrison}, {Gibbons}, {Goldstein}, {Gommers}, {Greco}, {Greenfield},
  {Groener}, {Grollier}, {Hagen}, {Hirst}, {Homeier}, {Horton}, {Hosseinzadeh},
  {Hu}, {Hunkeler}, {Ivezi{\'c}}, {Jain}, {Jenness}, {Kanarek}, {Kendrew},
  {Kern}, {Kerzendorf}, {Khvalko}, {King}, {Kirkby}, {Kulkarni}, {Kumar},
  {Lee}, {Lenz}, {Littlefair}, {Ma}, {Macleod}, {Mastropietro}, {McCully},
  {Montagnac}, {Morris}, {Mueller}, {Mumford}, {Muna}, {Murphy}, {Nelson},
  {Nguyen}, {Ninan}, {N{\"o}the}, {Ogaz}, {Oh}, {Parejko}, {Parley}, {Pascual},
  {Patil}, {Patil}, {Plunkett}, {Prochaska}, {Rastogi}, {Reddy Janga},
  {Sabater}, {Sakurikar}, {Seifert}, {Sherbert}, {Sherwood-Taylor}, {Shih},
  {Sick}, {Silbiger}, {Singanamalla}, {Singer}, {Sladen}, {Sooley},
  {Sornarajah}, {Streicher}, {Teuben}, {Thomas}, {Tremblay}, {Turner},
  {Terr{\'o}n}, {van Kerkwijk}, {de la Vega}, {Watkins}, {Weaver}, {Whitmore},
  {Woillez}, {Zabalza}, \& {Astropy
  Contributors}}]{Astropy-Collaboration:2018aa}
{Astropy Collaboration}, {Price-Whelan}, A.~M., {Sip{\H o}cz}, B.~M., {et~al.}
  2018, \aj, 156, 123, \dodoi{10.3847/1538-3881/aabc4f}

\bibitem[{{Basu}(1998)}]{Basu:1998aa}
{Basu}, S. 1998, \apj, 509, 229, \dodoi{10.1086/306494}

\bibitem[{{Bate}(2018)}]{Bate:2018aa}
{Bate}, M.~R. 2018, \mnras, 475, 5618, \dodoi{10.1093/mnras/sty169}

\bibitem[{{Bate} {et~al.}(2010){Bate}, {Lodato}, \& {Pringle}}]{Bate:2010aa}
{Bate}, M.~R., {Lodato}, G., \& {Pringle}, J.~E. 2010, \mnras, 401, 1505,
  \dodoi{10.1111/j.1365-2966.2009.15773.x}

\bibitem[{{Bjerkeli} {et~al.}(2016){Bjerkeli}, {J{\o}rgensen}, \&
  {Brinch}}]{Bjerkeli:2016ab}
{Bjerkeli}, P., {J{\o}rgensen}, J.~K., \& {Brinch}, C. 2016, \aap, 587, A145,
  \dodoi{10.1051/0004-6361/201527310}

\bibitem[{{Caselli} {et~al.}(2002){Caselli}, {Benson}, {Myers}, \&
  {Tafalla}}]{Caselli:2002aa}
{Caselli}, P., {Benson}, P.~J., {Myers}, P.~C., \& {Tafalla}, M. 2002, \apj,
  572, 238, \dodoi{10.1086/340195}

\bibitem[{{Chen} {et~al.}(2019){Chen}, {Storm}, {Li}, {Mundy}, {Frayer}, {Li},
  {Church}, {Friesen}, {Harris}, {Looney}, {Offner}, {Ostriker}, {Pineda},
  {Tobin}, \& {Chen}}]{Chen:2019ab}
{Chen}, C.-Y., {Storm}, S., {Li}, Z.-Y., {et~al.} 2019, \mnras, 490, 527,
  \dodoi{10.1093/mnras/stz2633}

\bibitem[{{Codella} {et~al.}(1999){Codella}, {Bachiller}, \&
  {Reipurth}}]{Codella:1999aa}
{Codella}, C., {Bachiller}, R., \& {Reipurth}, B. 1999, \aap, 343, 585

\bibitem[{{Dunham} \& {Vorobyov}(2012)}]{Dunham:2012aa}
{Dunham}, M.~M., \& {Vorobyov}, E.~I. 2012, \apj, 747, 52,
  \dodoi{10.1088/0004-637X/747/1/52}

\bibitem[{{Dunham} {et~al.}(2014){Dunham}, {Stutz}, {Allen}, {Evans},
  {Fischer}, {Megeath}, {Myers}, {Offner}, {Poteet}, {Tobin}, \&
  {Vorobyov}}]{Dunham:2014aa}
{Dunham}, M.~M., {Stutz}, A.~M., {Allen}, L.~E., {et~al.} 2014, Protostars and
  Planets VI, 195, \dodoi{10.2458/azu_uapress_9780816531240-ch009}

\bibitem[{{Dunham} {et~al.}(2015){Dunham}, {Allen}, {Evans},
  {Broekhoven-Fiene}, {Cieza}, {Di Francesco}, {Gutermuth}, {Harvey},
  {Hatchell}, {Heiderman}, {Huard}, {Johnstone}, {Kirk}, {Matthews}, {Miller},
  {Peterson}, \& {Young}}]{Dunham:2015aa}
{Dunham}, M.~M., {Allen}, L.~E., {Evans}, II, N.~J., {et~al.} 2015, \apjs, 220,
  11, \dodoi{10.1088/0067-0049/220/1/11}

\bibitem[{{Dutta} {et~al.}(2022){Dutta}, {Lee}, {Hirano}, {Liu}, {Johnstone},
  {Liu}, {Tatematsu}, {Goldsmith}, {Sahu}, {Evans}, {Sanhueza}, {Kwon}, {Qin},
  {Samal}, {Zhang}, {Kim}, {Shang}, {Lee}, {Moraghan}, {Jhan}, {Li}, {Lee},
  {Traficante}, {Juvela}, {Bronfman}, {Eden}, {Soam}, {He}, {Liu}, {Kuan},
  {Pelkonen}, {Luo}, {Yi}, \& {Hsu}}]{Dutta:2022aa}
{Dutta}, S., {Lee}, C.-F., {Hirano}, N., {et~al.} 2022, \apj, 931, 130,
  \dodoi{10.3847/1538-4357/ac67a1}

\bibitem[{{Evans} {et~al.}(2009){Evans}, {Dunham}, {J{\o}rgensen}, {Enoch},
  {Mer{\'{\i}}n}, {van Dishoeck}, {Alcal{\'a}}, {Myers}, {Stapelfeldt},
  {Huard}, {Allen}, {Harvey}, {van Kempen}, {Blake}, {Koerner}, {Mundy},
  {Padgett}, \& {Sargent}}]{Evans:2009aa}
{Evans}, II, N.~J., {Dunham}, M.~M., {J{\o}rgensen}, J.~K., {et~al.} 2009,
  \apjs, 181, 321, \dodoi{10.1088/0067-0049/181/2/321}

\bibitem[{{Foreman-Mackey} {et~al.}(2013){Foreman-Mackey}, {Hogg}, {Lang}, \&
  {Goodman}}]{Foreman-Mackey:2013aa}
{Foreman-Mackey}, D., {Hogg}, D.~W., {Lang}, D., \& {Goodman}, J. 2013, \pasp,
  125, 306, \dodoi{10.1086/670067}

\bibitem[{{Galametz} {et~al.}(2018){Galametz}, {Maury}, {Girart}, {Rao},
  {Zhang}, {Gaudel}, {Valdivia}, {Keto}, \& {Lai}}]{Galametz:2018aa}
{Galametz}, M., {Maury}, A., {Girart}, J.~M., {et~al.} 2018, \aap, 616, A139,
  \dodoi{10.1051/0004-6361/201833004}

\bibitem[{{Galametz} {et~al.}(2020){Galametz}, {Maury}, {Girart}, {Rao},
  {Zhang}, {Gaudel}, {Valdivia}, {Hennebelle}, {Cabedo-Soto}, {Keto}, \&
  {Lai}}]{Galametz:2020aa}
---. 2020, \aap, 644, A47, \dodoi{10.1051/0004-6361/202038854}

\bibitem[{{Gaudel} {et~al.}(2020){Gaudel}, {Maury}, {Belloche}, {Maret},
  {Andr{\'e}}, {Hennebelle}, {Galametz}, {Testi}, {Cabrit}, {Palmeirim},
  {Ladjelate}, {Codella}, \& {Podio}}]{Gaudel:2020aa}
{Gaudel}, M., {Maury}, A.~J., {Belloche}, A., {et~al.} 2020, \aap, 637, A92,
  \dodoi{10.1051/0004-6361/201936364}

\bibitem[{{Gupta} {et~al.}(2022){Gupta}, {Yen}, {Koch}, {Bastien}, {Bourke},
  {Chung}, {Hasegawa}, {Hull}, {Inutsuka}, {Kwon}, {Kwon}, {Lai}, {Lee}, {Lee},
  {Pattle}, {Qiu}, {Tahani}, {Tamura}, \& {Ward-Thompson}}]{Gupta:2022aa}
{Gupta}, A., {Yen}, H.-W., {Koch}, P., {et~al.} 2022, \apj, 930, 67,
  \dodoi{10.3847/1538-4357/ac63bc}

\bibitem[{{Hennebelle} {et~al.}(2020){Hennebelle}, {Commer{\c{c}}on}, {Lee}, \&
  {Charnoz}}]{Hennebelle:2020aa}
{Hennebelle}, P., {Commer{\c{c}}on}, B., {Lee}, Y.-N., \& {Charnoz}, S. 2020,
  \aap, 635, A67, \dodoi{10.1051/0004-6361/201936714}

\bibitem[{{Hirano} \& {Machida}(2019)}]{Hirano:2019aa}
{Hirano}, S., \& {Machida}, M.~N. 2019, \mnras, 485, 4667,
  \dodoi{10.1093/mnras/stz740}

\bibitem[{{Hull} {et~al.}(2014){Hull}, {Plambeck}, {Kwon}, {Bower},
  {Carpenter}, {Crutcher}, {Fiege}, {Franzmann}, {Hakobian}, {Heiles}, {Houde},
  {Hughes}, {Lamb}, {Looney}, {Marrone}, {Matthews}, {Pillai}, {Pound},
  {Rahman}, {Sandell}, {Stephens}, {Tobin}, {Vaillancourt}, {Volgenau}, \&
  {Wright}}]{Hull:2014aa}
{Hull}, C.~L.~H., {Plambeck}, R.~L., {Kwon}, W., {et~al.} 2014, \apjs, 213, 13,
  \dodoi{10.1088/0067-0049/213/1/13}

\bibitem[{Hunter(2007)}]{Hunter:2007aa}
Hunter, J.~D. 2007, Computing in Science \& Engineering, 9, 90,
  \dodoi{10.1109/MCSE.2007.55}

\bibitem[{{Johnstone} {et~al.}(2010){Johnstone}, {Rosolowsky}, {Tafalla}, \&
  {Kirk}}]{Johnstone:2010aa}
{Johnstone}, D., {Rosolowsky}, E., {Tafalla}, M., \& {Kirk}, H. 2010, \apj,
  711, 655, \dodoi{10.1088/0004-637X/711/2/655}

\bibitem[{{Joos} {et~al.}(2012){Joos}, {Hennebelle}, \& {Ciardi}}]{Joos:2012aa}
{Joos}, M., {Hennebelle}, P., \& {Ciardi}, A. 2012, \aap, 543, A128,
  \dodoi{10.1051/0004-6361/201118730}

\bibitem[{{Kauffmann} {et~al.}(2008){Kauffmann}, {Bertoldi}, {Bourke}, {Evans},
  \& {Lee}}]{Kauffmann:2008aa}
{Kauffmann}, J., {Bertoldi}, F., {Bourke}, T.~L., {Evans}, II, N.~J., \& {Lee},
  C.~W. 2008, \aap, 487, 993, \dodoi{10.1051/0004-6361:200809481}

\bibitem[{{Kenyon} {et~al.}(1990){Kenyon}, {Hartmann}, {Strom}, \&
  {Strom}}]{Kenyon:1990aa}
{Kenyon}, S.~J., {Hartmann}, L.~W., {Strom}, K.~M., \& {Strom}, S.~E. 1990,
  \aj, 99, 869, \dodoi{10.1086/115380}

\bibitem[{{Kirk} {et~al.}(2006){Kirk}, {Johnstone}, \& {Di
  Francesco}}]{Kirk:2006aa}
{Kirk}, H., {Johnstone}, D., \& {Di Francesco}, J. 2006, \apj, 646, 1009,
  \dodoi{10.1086/503193}

\bibitem[{{Kirk} {et~al.}(2007){Kirk}, {Johnstone}, \& {Tafalla}}]{Kirk:2007aa}
{Kirk}, H., {Johnstone}, D., \& {Tafalla}, M. 2007, \apj, 668, 1042,
  \dodoi{10.1086/521395}

\bibitem[{{Lee} {et~al.}(2000){Lee}, {Mundy}, {Reipurth}, {Ostriker}, \&
  {Stone}}]{Lee:2000aa}
{Lee}, C.-F., {Mundy}, L.~G., {Reipurth}, B., {Ostriker}, E.~C., \& {Stone},
  J.~M. 2000, \apj, 542, 925, \dodoi{10.1086/317056}

\bibitem[{{Lee} {et~al.}(2015){Lee}, {Dunham}, {Myers}, {Tobin}, {Kristensen},
  {Pineda}, {Vorobyov}, {Offner}, {Arce}, {Li}, {Bourke}, {J{\o}rgensen},
  {Goodman}, {Sadavoy}, {Chandler}, {Harris}, {Kratter}, {Looney}, {Melis},
  {Perez}, \& {Segura-Cox}}]{Lee:2015aa}
{Lee}, K.~I., {Dunham}, M.~M., {Myers}, P.~C., {et~al.} 2015, \apj, 814, 114,
  \dodoi{10.1088/0004-637X/814/2/114}

\bibitem[{{Lim} {et~al.}(2016){Lim}, {Hanawa}, {Yeung}, {Takakuwa},
  {Matsumoto}, \& {Saigo}}]{Lim:2016aa}
{Lim}, J., {Hanawa}, T., {Yeung}, P. K.~H., {et~al.} 2016, \apj, 831, 90,
  \dodoi{10.3847/0004-637X/831/1/90}

\bibitem[{{Machida} {et~al.}(2020){Machida}, {Hirano}, \&
  {Kitta}}]{Machida:2020aa}
{Machida}, M.~N., {Hirano}, S., \& {Kitta}, H. 2020, \mnras, 491, 2180,
  \dodoi{10.1093/mnras/stz3159}

\bibitem[{{Machida} \& {Matsumoto}(2011)}]{Machida:2011aa}
{Machida}, M.~N., \& {Matsumoto}, T. 2011, \mnras, 413, 2767,
  \dodoi{10.1111/j.1365-2966.2011.18349.x}

\bibitem[{{Machida} \& {Matsumoto}(2012)}]{Machida:2012aa}
---. 2012, \mnras, 421, 588, \dodoi{10.1111/j.1365-2966.2011.20336.x}

\bibitem[{{Matsumoto} {et~al.}(2017){Matsumoto}, {Machida}, \&
  {Inutsuka}}]{Matsumoto:2017aa}
{Matsumoto}, T., {Machida}, M.~N., \& {Inutsuka}, S.-i. 2017, \apj, 839, 69,
  \dodoi{10.3847/1538-4357/aa6a1c}

\bibitem[{{Matsumoto} \& {Tomisaka}(2004)}]{Matsumoto:2004aa}
{Matsumoto}, T., \& {Tomisaka}, K. 2004, \apj, 616, 266, \dodoi{10.1086/424897}

\bibitem[{{McMullin} {et~al.}(2007){McMullin}, {Waters}, {Schiebel}, {Young},
  \& {Golap}}]{McMullin:2007aa}
{McMullin}, J.~P., {Waters}, B., {Schiebel}, D., {Young}, W., \& {Golap}, K.
  2007, in Astronomical Society of the Pacific Conference Series, Vol. 376,
  Astronomical Data Analysis Software and Systems XVI, ed. R.~A. {Shaw},
  F.~{Hill}, \& D.~J. {Bell}, 127

\bibitem[{{Mellon} \& {Li}(2008)}]{Mellon:2008aa}
{Mellon}, R.~R., \& {Li}, Z.-Y. 2008, \apj, 681, 1356, \dodoi{10.1086/587542}

\bibitem[{{Offner} \& {Chaban}(2017)}]{Offner:2017aa}
{Offner}, S. S.~R., \& {Chaban}, J. 2017, \apj, 847, 104,
  \dodoi{10.3847/1538-4357/aa8996}

\bibitem[{{Offner} {et~al.}(2011){Offner}, {Lee}, {Goodman}, \&
  {Arce}}]{Offner:2011aa}
{Offner}, S. S.~R., {Lee}, E.~J., {Goodman}, A.~A., \& {Arce}, H. 2011, \apj,
  743, 91, \dodoi{10.1088/0004-637X/743/1/91}

\bibitem[{{Ohashi} {et~al.}(2023){Ohashi}, {Tobin}, {J{\o}rgensen}, {Takakuwa},
  {Sheehan}, {Aikawa}, {Li}, {Looney}, {Williams}, {Aso}, {Sharma}, {Choi},
  {Yamato}, {Lee}, {Tomida}, {Yen}, {Encalada}, {Flores}, {Gavino}, {Kido},
  {Han}, {Lin}, {Narayanan}, {Phuong}, {Santamar{\'\i}a-Miranda}, {Thieme},
  {van't Hoff}, {de Gregorio-Monsalvo}, {Koch}, {Kwon}, {Lai}, {Lee},
  {Plunkett}, {Saigo}, {Hirano}, {Lam}, \& {Mori}}]{Ohashi:2023aa}
{Ohashi}, N., {Tobin}, J.~J., {J{\o}rgensen}, J.~K., {et~al.} 2023, \apj, 951,
  8, \dodoi{10.3847/1538-4357/acd384}

\bibitem[{{Okoda} {et~al.}(2018){Okoda}, {Oya}, {Sakai}, {Watanabe},
  {J{\o}rgensen}, {Van Dishoeck}, \& {Yamamoto}}]{Okoda:2018aa}
{Okoda}, Y., {Oya}, Y., {Sakai}, N., {et~al.} 2018, \apjl, 864, L25,
  \dodoi{10.3847/2041-8213/aad8ba}

\bibitem[{{Okoda} {et~al.}(2021){Okoda}, {Oya}, {Francis}, {Johnstone},
  {Inutsuka}, {Ceccarelli}, {Codella}, {Chandler}, {Sakai}, {Aikawa}, {Alves},
  {Balucani}, {Bianchi}, {Bouvier}, {Caselli}, {Caux}, {Charnley}, {Choudhury},
  {De Simone}, {Dulieu}, {Dur{\'a}n}, {Evans}, {Favre}, {Fedele}, {Feng},
  {Fontani}, {Hama}, {Hanawa}, {Herbst}, {Hirota}, {Imai}, {Isella},
  {J{\'\i}menez-Serra}, {Kahane}, {Lefloch}, {Loinard}, {L{\'o}pez-Sepulcre},
  {Maud}, {Maureira}, {Menard}, {Mercimek}, {Miotello}, {Moellenbrock}, {Mori},
  {Murillo}, {Nakatani}, {Nomura}, {Oba}, {O'Donoghue}, {Ohashi},
  {Ospina-Zamudio}, {Pineda}, {Podio}, {Rimola}, {Sakai}, {Segura-Cox},
  {Shirley}, {Svoboda}, {Taquet}, {Testi}, {Vastel}, {Viti}, {Watanabe},
  {Watanabe}, {Witzel}, {Xue}, {Zhang}, {Zhao}, \& {Yamamoto}}]{Okoda:2021aa}
{Okoda}, Y., {Oya}, Y., {Francis}, L., {et~al.} 2021, \apj, 910, 11,
  \dodoi{10.3847/1538-4357/abddb1}

\bibitem[{Oliphant(2006)}]{Oliphant:2006aa}
Oliphant, T.~E. 2006, A guide to {NumPy}, USA: Trelgol Publishing

\bibitem[{{Ossenkopf} \& {Henning}(1994)}]{Ossenkopf:1994aa}
{Ossenkopf}, V., \& {Henning}, T. 1994, \aap, 291, 943

\bibitem[{{Oya} {et~al.}(2014){Oya}, {Sakai}, {Sakai}, {Watanabe}, {Hirota},
  {Lindberg}, {Bisschop}, {J{\o}rgensen}, {van Dishoeck}, \&
  {Yamamoto}}]{Oya:2014aa}
{Oya}, Y., {Sakai}, N., {Sakai}, T., {et~al.} 2014, \apj, 795, 152,
  \dodoi{10.1088/0004-637X/795/2/152}

\bibitem[{{Redaelli} {et~al.}(2019){Redaelli}, {Alves}, {Santos}, \&
  {Caselli}}]{Redaelli:2019aa}
{Redaelli}, E., {Alves}, F.~O., {Santos}, F.~P., \& {Caselli}, P. 2019, \aap,
  631, A154, \dodoi{10.1051/0004-6361/201936271}

\bibitem[{{Sai} {et~al.}(2023){Sai}, {Ohashi}, {Yen}, {Maury}, \&
  {Maret}}]{Sai:2023aa}
{Sai}, J. I.~C., {Ohashi}, N., {Yen}, H.-W., {Maury}, A.~J., \& {Maret}, S.
  2023, \apj, 944, 222, \dodoi{10.3847/1538-4357/acb3bd}

\bibitem[{{Santamar{\'\i}a-Miranda} {et~al.}(2021){Santamar{\'\i}a-Miranda},
  {de Gregorio-Monsalvo}, {Plunkett}, {Hu{\'e}lamo}, {L{\'o}pez}, {Ribas},
  {Schreiber}, {Mu{\v{z}}i{\'c}}, {Palau}, {Knee}, {Bayo}, {Comer{\'o}n}, \&
  {Hales}}]{Santamaria-Miranda:2021aa}
{Santamar{\'\i}a-Miranda}, A., {de Gregorio-Monsalvo}, I., {Plunkett}, A.~L.,
  {et~al.} 2021, \aap, 646, A10, \dodoi{10.1051/0004-6361/202039419}

\bibitem[{{Sato} {et~al.}(2023){Sato}, {Tokuda}, {Machida}, {Tachihara},
  {Harada}, {Yamasaki}, {Hirano}, {Onishi}, \& {Matsushita}}]{Sato:2023aa}
{Sato}, A., {Tokuda}, K., {Machida}, M.~N., {et~al.} 2023, \apj, 958, 102,
  \dodoi{10.3847/1538-4357/ad0132}

\bibitem[{{Seifried} {et~al.}(2013){Seifried}, {Banerjee}, {Pudritz}, \&
  {Klessen}}]{Seifried:2013aa}
{Seifried}, D., {Banerjee}, R., {Pudritz}, R.~E., \& {Klessen}, R.~S. 2013,
  \mnras, 432, 3320, \dodoi{10.1093/mnras/stt682}

\bibitem[{{Shu} {et~al.}(1991){Shu}, {Ruden}, {Lada}, \& {Lizano}}]{Shu:1991aa}
{Shu}, F.~H., {Ruden}, S.~P., {Lada}, C.~J., \& {Lizano}, S. 1991, \apjl, 370,
  L31, \dodoi{10.1086/185970}

\bibitem[{{Takahashi} {et~al.}(2016){Takahashi}, {Tomida}, {Machida}, \&
  {Inutsuka}}]{Takahashi:2016aa}
{Takahashi}, S.~Z., {Tomida}, K., {Machida}, M.~N., \& {Inutsuka}, S.-i. 2016,
  \mnras, 463, 1390, \dodoi{10.1093/mnras/stw1994}

\bibitem[{{Terebey} {et~al.}(1984){Terebey}, {Shu}, \&
  {Cassen}}]{Terebey:1984aa}
{Terebey}, S., {Shu}, F.~H., \& {Cassen}, P. 1984, \apj, 286, 529,
  \dodoi{10.1086/162628}

\bibitem[{{Thieme} {et~al.}(2023){Thieme}, {Lai}, {Ohashi}, {Tobin},
  {J{\o}rgensen}, {Sai}, {Aso}, {Williams}, {Yamato}, {Aikawa}, {de
  Gregorio-Monsalvo}, {Han}, {Kwon}, {Lee}, {Lee}, {Li}, {Lin}, {Looney},
  {Narayanan}, {Phuong}, {Plunkett}, {Santamar{\'\i}a-Miranda}, {Sharma},
  {Takakuwa}, \& {Yen}}]{Thieme:2023aa}
{Thieme}, T.~J., {Lai}, S.-P., {Ohashi}, N., {et~al.} 2023, \apj, 958, 60,
  \dodoi{10.3847/1538-4357/ad003a}

\bibitem[{{Tobin} {et~al.}(2016){Tobin}, {Kratter}, {Persson}, {Looney},
  {Dunham}, {Segura-Cox}, {Li}, {Chandler}, {Sadavoy}, {Harris}, {Melis}, \&
  {P{\'e}rez}}]{Tobin:2016aa}
{Tobin}, J.~J., {Kratter}, K.~M., {Persson}, M.~V., {et~al.} 2016, \nat, 538,
  483, \dodoi{10.1038/nature20094}

\bibitem[{{Tobin} {et~al.}(2020){Tobin}, {Sheehan}, {Megeath},
  {D{\'\i}az-Rodr{\'\i}guez}, {Offner}, {Murillo}, {van 't Hoff}, {van
  Dishoeck}, {Osorio}, {Anglada}, {Furlan}, {Stutz}, {Reynolds}, {Karnath},
  {Fischer}, {Persson}, {Looney}, {Li}, {Stephens}, {Chandler}, {Cox},
  {Dunham}, {Tychoniec}, {Kama}, {Kratter}, {Kounkel}, {Mazur}, {Maud},
  {Patel}, {Perez}, {Sadavoy}, {Segura-Cox}, {Sharma}, {Stephenson}, {Watson},
  \& {Wyrowski}}]{Tobin:2020aa}
{Tobin}, J.~J., {Sheehan}, P.~D., {Megeath}, S.~T., {et~al.} 2020, \apj, 890,
  130, \dodoi{10.3847/1538-4357/ab6f64}

\bibitem[{{Tomida} {et~al.}(2015){Tomida}, {Okuzumi}, \&
  {Machida}}]{Tomida:2015aa}
{Tomida}, K., {Okuzumi}, S., \& {Machida}, M.~N. 2015, \apj, 801, 117,
  \dodoi{10.1088/0004-637X/801/2/117}

\bibitem[{{van der Walt} {et~al.}(2011){van der Walt}, {Colbert}, \&
  {Varoquaux}}]{van-der-Walt:2011aa}
{van der Walt}, S., {Colbert}, S.~C., \& {Varoquaux}, G. 2011, Computing in
  Science and Engineering, 13, 22, \dodoi{10.1109/MCSE.2011.37}

\bibitem[{{van Kempen} {et~al.}(2009){van Kempen}, {van Dishoeck},
  {Hogerheijde}, \& {G{\"u}sten}}]{van-Kempen:2009ac}
{van Kempen}, T.~A., {van Dishoeck}, E.~F., {Hogerheijde}, M.~R., \&
  {G{\"u}sten}, R. 2009, \aap, 508, 259, \dodoi{10.1051/0004-6361/200811099}

\bibitem[{{Virtanen} {et~al.}(2020){Virtanen}, {Gommers}, {Oliphant},
  {Haberland}, {Reddy}, {Cournapeau}, {Burovski}, {Peterson}, {Weckesser},
  {Bright}, {van der Walt}, {Brett}, {Wilson}, {Jarrod Millman}, {Mayorov},
  {Nelson}, {Jones}, {Kern}, {Larson}, {Carey}, {Polat}, {Feng}, {Moore}, {Vand
  erPlas}, {Laxalde}, {Perktold}, {Cimrman}, {Henriksen}, {Quintero}, {Harris},
  {Archibald}, {Ribeiro}, {Pedregosa}, {van Mulbregt}, \&
  {Contributors}}]{Virtanen:2020aa}
{Virtanen}, P., {Gommers}, R., {Oliphant}, T.~E., {et~al.} 2020, Nature
  Methods, 17, 261, \dodoi{https://doi.org/10.1038/s41592-019-0686-2}

\bibitem[{{Ward-Thompson} {et~al.}(2007){Ward-Thompson}, {Andr{\'e}},
  {Crutcher}, {Johnstone}, {Onishi}, \& {Wilson}}]{Ward-Thompson:2007ab}
{Ward-Thompson}, D., {Andr{\'e}}, P., {Crutcher}, R., {et~al.} 2007, Protostars
  and Planets V, 33

\bibitem[{{Williams} \& {Cieza}(2011)}]{Williams:2011aa}
{Williams}, J.~P., \& {Cieza}, L.~A. 2011, \araa, 49, 67,
  \dodoi{10.1146/annurev-astro-081710-102548}

\bibitem[{{Yen} {et~al.}(2017){Yen}, {Koch}, {Takakuwa}, {Krasnopolsky},
  {Ohashi}, \& {Aso}}]{Yen:2017aa}
{Yen}, H.-W., {Koch}, P.~M., {Takakuwa}, S., {et~al.} 2017, \apj, 834, 178,
  \dodoi{10.3847/1538-4357/834/2/178}

\bibitem[{{Yen} \& {the eDisk team}(2024)}]{Yen:inprep}
{Yen}, H.-W., \& {the eDisk team}. 2024, \apj, submitted

\bibitem[{{Yen} {et~al.}(2021{\natexlab{a}}){Yen}, {Zhao}, {Koch}, \&
  {Gupta}}]{Yen:2021ab}
{Yen}, H.-W., {Zhao}, B., {Koch}, P.~M., \& {Gupta}, A. 2021{\natexlab{a}},
  \apj, 916, 97, \dodoi{10.3847/1538-4357/ac0723}

\bibitem[{{Yen} {et~al.}(2021{\natexlab{b}}){Yen}, {Koch}, {Hull},
  {Ward-Thompson}, {Bastien}, {Hasegawa}, {Kwon}, {Lai}, {Qiu}, {Ching},
  {Chung}, {Coud{\'e}}, {Di Francesco}, {Diep}, {Doi}, {Eswaraiah}, {Falle},
  {Fuller}, {Furuya}, {Han}, {Hatchell}, {Houde}, {Inutsuka}, {Johnstone},
  {Kang}, {Kang}, {Kim}, {Kirchschlager}, {Kwon}, {Lee}, {Lee}, {Liu}, {Liu},
  {Lyo}, {Ohashi}, {Onaka}, {Pattle}, {Sadavoy}, {Saito}, {Shinnaga}, {Soam},
  {Tahani}, {Tamura}, {Tang}, {Tang}, \& {Zhang}}]{Yen:2021aa}
{Yen}, H.-W., {Koch}, P.~M., {Hull}, C. L.~H., {et~al.} 2021{\natexlab{b}},
  \apj, 907, 33, \dodoi{10.3847/1538-4357/abca99}

\bibitem[{{Y{\i}ld{\i}z} {et~al.}(2012){Y{\i}ld{\i}z}, {Kristensen}, {van
  Dishoeck}, {Belloche}, {van Kempen}, {Hogerheijde}, {G{\"u}sten}, \& {van der
  Marel}}]{Yildiz:2012aa}
{Y{\i}ld{\i}z}, U.~A., {Kristensen}, L.~E., {van Dishoeck}, E.~F., {et~al.}
  2012, \aap, 542, A86, \dodoi{10.1051/0004-6361/201118368}

\bibitem[{{Y{\i}ld{\i}z} {et~al.}(2013){Y{\i}ld{\i}z}, {Kristensen}, {van
  Dishoeck}, {San Jos{\'e}-Garc{\'\i}a}, {Karska}, {Harsono}, {Tafalla},
  {Fuente}, {Visser}, {J{\o}rgensen}, \& {Hogerheijde}}]{Yildiz:2013aa}
---. 2013, \aap, 556, A89, \dodoi{10.1051/0004-6361/201220849}

\bibitem[{{Y{\i}ld{\i}z} {et~al.}(2015){Y{\i}ld{\i}z}, {Kristensen}, {van
  Dishoeck}, {Hogerheijde}, {Karska}, {Belloche}, {Endo}, {Frieswijk},
  {G{\"u}sten}, {van Kempen}, {Leurini}, {Nagy}, {P{\'e}rez-Beaupuits},
  {Risacher}, {van der Marel}, {van Weeren}, \& {Wyrowski}}]{Yildiz:2015aa}
---. 2015, \aap, 576, A109, \dodoi{10.1051/0004-6361/201424538}

\bibitem[{{Zucker} {et~al.}(2020){Zucker}, {Speagle}, {Schlafly}, {Green},
  {Finkbeiner}, {Goodman}, \& {Alves}}]{Zucker:2020aa}
{Zucker}, C., {Speagle}, J.~S., {Schlafly}, E.~F., {et~al.} 2020, \aap, 633,
  A51, \dodoi{10.1051/0004-6361/201936145}

\bibitem[{{Zucker} {et~al.}(2022){Zucker}, {Goodman}, {Alves}, {Bialy},
  {Foley}, {Speagle}, {Gro{\ss}schedl}, {Finkbeiner}, {Burkert}, {Khimey}, \&
  {Swiggum}}]{Zucker:2022aa}
{Zucker}, C., {Goodman}, A.~A., {Alves}, J., {et~al.} 2022, arXiv e-prints,
  arXiv:2201.05124.
\newblock \doarXiv{2201.05124}

\end{thebibliography}
\bibliographystyle{aasjournal}



\end{document}